\documentclass[journal]{IEEEtran}
\usepackage{graphicx}
\usepackage{amsmath}
\begin{document}
\title{Dust-ion-acoustic solitons in an ion-beam-driven dusty magnetoplasma with adiabatic and nonadiabatic dust charge variations}

\author{Num Prasad Acharya, Suresh Basnet, Amar Prasad Misra
        and Raju Khanal ~\IEEEmembership{Member,~IEEE}
\thanks{N. P. Acharya is with the Central Department of Physics, Tribhuvan University, Kirtipur, Kathmandu 44613, Nepal, and also at Department of Physics, Mahendra Multiple Campus, Tribhuvan University, Ghorahi 22415, Dang, Nepal} 
\thanks{S. Basnet is with the Asia Pacific Center for Theoretical Physics, Pohang, Gyeongbuk 37673, Republic of Korea. e-mail: (suresh.basnet@apctp.org).}
\thanks{A. P. Misra is with the Department of Mathematics, Siksha Bhavana, Visva-Bharati University, Santiniketan-731 235, India. e-mail: (apmisra@visva-bharati.ac.in).}
\thanks{R. Khanal is with the Central Department of Physics, Tribhuvan University, Kirtipur, Kathmandu 44613, Nepal.}}
\maketitle
\begin{abstract}
We study the characteristics of small-amplitude nonlinear dust-ion-acoustic (DIA) solitary waves in active magnetized positive-ion-beam-driven dusty plasmas with the effects of nonadiabatic and adiabatic dust charge variations. In the model, we consider the ion-neutral collision and thereby consider the collision enhanced ion current to the dust-charging process and dust charge fluctuations. We show that the streaming of the positive-ion beam significantly affects the dust-charging process in which the dust charge number decreases (increases) with an increased beam velocity (number density). Using the standard reductive perturbation technique, we derive the evolution equations in the form of Korteweg-de Vries (KdV) equations for DIA solitary waves for two different cases: nonadiabatic and adiabatic dust charge variations. We study the effect of positive ion beam, dust charge variation, magnetic field, ion creation, and ion-neutral collision enhanced current on the wave characteristics. We find that the soliton energy decays with time and is affected by the beam velocity. Also, the solitary waves get damped by the effects of ion creation, ion loss, ion-neutral collision enhanced current, and dust charge variation. Although the ion beam does not change the polarity of solitary waves in the case of adiabatic dust charge variation, a transition from rarefactive to compressive solitary waves occurs in the presence of an ion beam with nonadiabatic dust charge variation.     
\end{abstract}

\IEEEpeerreviewmaketitle

\section{Introduction}
\IEEEPARstart
{D}{usty} plasmas or complex plasmas have received a great research interest in exploring the novel characteristics of dusty plasmas, which comprise additional positively or negatively charged micrometer-sized dust grains in an ordinary ion-electron plasma \cite{el2007sagdeev}, also known as multi-component plasmas. The dust particulates acquire an electric charge by collecting local plasma electrons and positive ions. In most cases, dust grains are negatively charged due to a higher rate of electron flux residing on the surface of dust grains. However, there are various phenomena (such as electron emission phenomena, e.g., photoelectric emission, thermionic emission, and secondary electron emission from the dust grains) occur in dusty plasma systems in which dust grains are positively charged \cite{cui1994fluctuations}. The size of dust particles affects the magnitude of the charge on dust ($q_{\textrm{d}}$), i.e. larger the size, the higher will be the charge ($q_{\mathrm{d}} = C\phi_{\mathrm{d}}$ with $C = 4\pi\epsilon_0 r_{\mathrm{d}}$ denoting the capacitance of dust grains, and $r_{\mathrm{d}}$ is the dust-grain radius). The presence of charged dust grains is considered to be crucial in many laboratory plasmas \cite{kersten2001micro} \cite{wang2001ionization} as well as space and astrophysical plasmas \cite{goertz1989dusty}. Not only do these charged dust grains modify the linear and nonlinear properties and quasineutrality condition of an ordinary plasma but also introduce new types of waves in dusty plasmas \cite{tribeche2009effect}. 
\par
The fluctuation in background ion and electron currents flowing on the surface of dust grains causes the dust charge to diversify. The two limiting cases of dust charge variations are of interest: adiabatic and non-adiabatic processes. Thus, the dust charge becomes a new dynamic variable, and its effects on the characteristics of collective plasma motion of plasma are crucial in determining the various waves in dusty plasmas \cite{mamun2002role} \cite{mowafy2008effect} \cite{alinejad2011effects}. In the linear regime, the nonadiabatic dust charge variation causes damping of wave eigenmodes \cite{verheest1997dust}. However, in nonlinear regimes, the nonadiabatic dust charge variation causes an anomalous dissipation in dusty plasmas, which may be responsible for the generation of shock waves in dusty plasmas \cite{bai2000dust} \cite{gupta2001effect}. 
\par 
Different fundamental modes can be excited and their propagation characteristics also differ in dusty plasmas depending on whether the charged dust grains are mobile or static \cite{adhikary2010ion}. In this context, dust-acoustic waves (DAWs) \cite{rao1990dust} and dust-ion-acoustic waves (DIAWs) \cite{shukla1992dust} are the two fundamental wave modes observed in dusty plasmas. The DIAWs can be damped due to dust charge fluctuations induced by the energy exchange during the dust charging process \cite{ma1994self}. Besides the dust charge fluctuation, the presence of negative ions, ion generation, and loss terms, and the type of electron distribution determine the instability or damping of DIAWs \cite{vladimirov2003ion} \cite{chen2019effect}. In the nonlinear regime, the nonlinear coherent structures such as dust-ion acoustic solitary waves (DIASWs) \cite{mamun2002role}, ion-acoustic solitary waves (IASWs)  \cite{goswami2018study} \cite{deka2012characteristics}, and electron-acoustic solitary waves (EASWs) \cite{devanandhan2011electron} can be formed in different plasma environments.
In the past few years, the effects of electron distribution, oblique magnetic field, the concentration of dust grains, ion-neutral collision, ion-dust collision, ion source, and ion loss on the propagation characteristics of finite-amplitude DIASWs have been studied owing to their relevance in laboratory and space plasmas \cite{choi2007dust} \cite{shalaby2009stability} \cite{mayout2016effects} \cite{tamang2018solitary} \cite{dehingia2023propagation}.
\par 
Positive ion beams in the dust-charge dynamics can play a decisive role in transferring charges through impacts. Typically, they deposit a positive charge to the dust grains upon impacting them, thereby increasing the electric field potential at the grain surface. Such a charging process significantly influences the dynamics of dust grains, as well as various wave modes and nonlinear coherent structures in dusty plasmas \cite{okutsu1978amplification} \cite{nejoh1999nonlinear}  \cite{pavluu2006ion} \cite{adhikary2010ion}. For ion beam-driven laboratory dusty plasmas, it is indispensable to investigate plasma-wall interactions such as surface cleaning, thin film deposition, chemical vapor deposition, micro-fabrication, Coulomb crystals (to explore the properties of strongly coupled plasmas), and surface modification of materials \cite{kersten2001micro}, \cite{kersten2006interaction} \cite{bonitz2012magnetized}. Besides applications in laboratory plasmas, the ion beam-driven dusty plasmas are equally significant in space and astrophysical environments (e.g., auroral zone of the upper atmosphere, planetary rings, etc.) \cite{kersten2006interaction}. Theoretically, the effect of the ion beam on large amplitude DAWs with temporal evolution of the dust-grain charge was studied. It has been shown that the dust charge number can increase with an increase in the ion-beam temperature and plasma number density \cite{nejoh1999nonlinear}. However, the increment of the dust charge modifies the wave phase speed. The arbitrary streaming of ion beams in dusty plasmas can result in both compressive and rarefactive solitons along with double layers. In this context, a critical beam velocity was obtained below which no solitons would exist \cite{el2003dust}.
\par 
 The ion creation and an ion loss in dusty plasmas can also modify the nonlinear features of DIAWs due to weak dissipation. While the ion creation causes ion-acoustic waves to be unstable, the ion loss and dust charge fluctuation can play significant roles in determining the ionization instability \cite{ghosh2007weakly}. The presence of ion-neutral and ion-dust collisions in dusty plasmas with an ionization effect modifies the nonlinear propagation of DIASWs. In the absence of ion-dust and ion-neutral collisions, the dominant ionization over the ion loss process can lead to instability of DIAWs. However, if collision effects dominate over the ionization, DIAWs get damped \cite{shalaby2010dust}.
 
Adhikary \textit{et al.} studied the nonlinear propagation of DIAWs in the presence of positive ion beams in dusty plasmas using the reductive perturbation technique. They reported critical values of the ion-beam speed, above and below which the DIA solitons do not exist \cite{adhikary2010ion}. Using Sagdeev’s potential approach, Chatterjee \textit{et al.} \cite{chatterjee2012dust} derived the conditions for the existence of DASWs in unmagnetized homogeneous ion-beam driven dusty plasmas with dust charge fluctuations. They showed the existence of a critical value of the dust streaming velocity beyond which the solitary waves would not exist. Experimentally, the nonlinear propagation of DIASWs in an unmagnetized dusty plasma under the influence of ion beam and charged dust particulates has been studied by Deka \textit{et al.} They showed that the solitary waves appear both compressive and rarefactive types associated with the fast and slow beam modes \cite{deka2012characteristics}.

Atteya \textit{et al.} \cite{atteya2018dust} studied obliquely propagating DIASWs in a magnetized electronegative dusty plasma by using the reductive perturbation method. They observed the fast and slow modes in the linear regime and studied the dynamics of DIASWs by the influences of the kappa distribution of electrons, the temperature of ions, the obliqueness of wave propagation, and the magnetic field. Recently, the theoretical analysis of nonlinear DIAWs for magnetized dusty plasmas presented to examine the effect of magnetic field, dust charge variation, obliqueness of wave propagation, and superthermal index of electron distribution on the wave dynamics \cite{abdus2022higher} \cite{salam2024effects}. To incorporate the effects of higher-order nonlinear terms on the wave's structure, the modified Korteweg de-Vries (mKdV) equation was derived using the reductive perturbation technique. It was shown that the inclusion of higher-order terms, magnetic field, dust density, and adiabatic dust charge variation considerably modify the wave phase speed and profiles of nonlinear coherent structures  \cite{abdus2022higher} \cite{salam2024effects}.  
\par 
All of the above experimental and theoretical works extensively discussed the effects of ion-beam on dust charging processes and wave characteristics either in magnetized or unmagnetized plasmas but in the absence of source and sink terms, ion-neutral collisions, and dust charge fluctuations with collision enhanced current. Therefore, a proper theoretical investigation of DIAWs in the presence of all these effects in ion-beam-driven dusty plasmas is still lacking and needs special attention. 

In this work, we aim to study the dust-ion-acoustic solitary waves in active magnetized ion-beam-driven dusty plasmas including the collision enhanced ion current during the dust charging process. To this end, we numerically calculate the equilibrium dust charge number in the presence of an ion-neutral collision enhanced current by the Newton-Raphson method. We present the effects of the obliqueness of wave propagation about the magnetic field, ion-beam streaming velocity, beam density, ion creation, ion loss, and dust charge fluctuation on the profiles of DIA solitons for two different cases: nonadiabatic and adiabatic dust charge variations. 
\par 
The manuscript is organized in the following way. We present the basic equations for electrons and ion fluids and the dust-charging equation in Sec. \ref{sec-basic}. Section \ref{sec-nonl} presents the nonlinear evolution equations for DIA solitary waves and the soliton solutions in two cases of nonadiabatic and adiabatic dust charge variations. We numerically analyze the soliton solutions by the parameters relevant for laboratory dusty plasmas in Sec. \ref{sec-resul}. Finally, we leave Sec. \ref{sec-summ} to summarize the our results and conclude.
\section{MODEL AND GOVERNING EQUATIONS}
\label{sec-basic}
We consider a collisional magnetized dusty plasma (static magnetic field), which consists of inertial positive ions, Maxwellian electrons, and positive ion beam, and immobile negatively charged dust grains.  We assume that the ion beam has a streaming velocity, $\text{v}_{\mathrm{b}0}$ smaller than its thermal velocity and dust grains are massive compared to positive ions and electrons such that their time scale of oscillation is longer than the electron and ion plasma periods. 
\par 
At equilibrium, the  macroscopic charge neutrality condition reads
\begin{equation}
en_{\textrm{i}0} + en_{\textrm{b}0} = en_{\textrm{e}0}-q_{\textrm{d}0}n_{\textrm{d}0},
\end{equation}
where $e$ is the elementary charge, $q_{\mathrm{d}0}$ is the unperturbed dust charge, and $n_{\mathrm{s}0}$ is the unperturbed number density of particle species-$s$ with $\mathrm{s} = `\mathrm{i}$' for positive ions, `$\mathrm{e}$' for electrons, `$\mathrm{b}$' for streaming positive ion beams, and `$\mathrm{d}$' for negatively charged dust grains. 
For simplicity, we assume the wave propagation along the $x$-direction and the external uniform magnetic field $\mathbf{B}$ in the $xz$-plane, which makes an angle $\theta$ with the wave propagation vector $\textbf{k}=k\hat{x}$ such that $\mathbf{B}=(B\cos\theta,0,B\sin\theta)$.

Typically, when electrons reach dust grain surfaces faster than positive ions, the dust grains become negatively charged. The total current flowing on the dust-grain surface is the sum of electron ($I_{\textrm{e}}$) and total ion ($I_{\textrm{Ti}}$) currents (where $I_{\textrm{Ti}}$ consists of the positive ion current $I_{\textrm{i}}$, collision enhanced current $I_{\textrm{coll}}$), and positive ion-beam current $I_{\textrm{b}}$. Thus, the temporal evolution of the dust charge is
\begin{equation}
\frac{dq_{\textrm{d}}}{dt} = I_{\textrm{Ti}} + I_{\textrm{b}} + I_{\textrm{e}}.
\label{eqn_dust_charge}
\end{equation}
Using the orbital motion limited (OML) theory \cite{khrapak2005particle} \cite{barkan1994charging} \cite{shukla2001survey} \cite{mamun2003charging}, the ion, beam, and electron currents are obtained as
\begin{equation}
\begin{split}
I_{\mathrm{Ti}} = &\pi r^{2}_{\mathrm{d}}en_{\mathrm{i}}\sqrt{\frac{8T_{\mathrm{i}}}{\pi m_{\mathrm{i}}}}\left[1- \frac{Z}{\sigma_{\textrm{i}}}\left(\Delta Q -1\right)\right.\\
 &\left.+\frac{0.4\lambda_{\mathrm{D}}}{\lambda_{\mathrm{in}}} \frac{Z^{2}\left(\Delta Q-1\right)^{2}}{\sigma_{\textrm{i}}^{2}}\right],
 \end{split}
\label{eqn_ion_current}
\end{equation}
\begin{equation}
I_{\textrm{b}} =  \pi r_{\textrm{d}}^{2}en_{\textrm{b}0}\text{v}_{\textrm{b}0}\left[1-\frac{2m_{\textrm{i}}Z(\Delta Q -1)}{m_{\mathrm{b}}u_{\textrm{b}0}^{2}}\right]\exp\left(-\frac{\psi}{\sigma_{\textrm{b}}}\right)
\label{eqn_beam_current}
\end{equation}
and
\begin{equation}
I_{\textrm{e}} = -\pi r_{\textrm{d}}^{2} e n_{\textrm{e}0} \left( \frac{8T_{\textrm{e}}}{\pi m_{\textrm{e}}} \right)^{\frac{1}{2}}\exp\left( {Z(\Delta Q-1)}+\psi\right),
\label{eqn_electron_current}
\end{equation}
where $u_{\textrm{b}0} = \text{v}_{\textrm{b}0} / C_{\textrm{s}}$ is the normalized positive ion beam velocity with $C_{\textrm{s}} = \sqrt{T_{\textrm{e}}/m_{\textrm{i}}}$ denoting the ion-acoustic speed (in which the electron temperature $T_e$ is in energy units), $\sigma_{\textrm{i}} = T_{\textrm{i}} / T_{\textrm{e}}$ is the ratio of ion to electron temperatures, $\sigma_{\textrm{b}} = T_{\textrm{b}} /T_{\textrm{e}}$ is the ratio of ion-beam to electron temperatures, $Z = Z_{\textrm{d}0}e^{2}/4\pi \epsilon_{0}r_{\textrm{d}}T_{\textrm{e}}$, $\lambda_{\textrm{in}} = 1/n_{\textrm{n}}\sigma_{\textrm{s}}$ is the mean free path for ion-neutral collisions with the neutral gas density $n_{\textrm{n}}$ and collision cross section $\sigma_{\textrm{s}}$, $\Delta Q = q_{\textrm{d1}}/Z_{\mathrm{d}0}e$ is the dust charge fluctuation   normalized by the equilibrium dust charge $e Z_{\textrm{d}0}$, and $\psi = e\phi/T_{\textrm{e}}$ is normalized electrostatic potential. We note that in presence of the streaming ion beam, the dust screening length $\lambda_{\textrm{D}}$ gets modified and is given by 
\begin{equation}
\lambda_{\textrm{D}} = \lambda_{\textrm{De}} \left[\delta_{\textrm{e}}+\frac{1}{\sigma_{\textrm{i}}}+\frac{m_{\textrm{i}}\delta_{\textrm{b}}}{m_{\textrm{b}}\sigma_{\textrm{b}}u_{\textrm{b}0}^{2}}\right]^{-1/2}
\label{eqn_screening_length}
\end{equation}
with $\delta_{\textrm{e}} = n_{\textrm{e}0}/n_{\textrm{i}0}$  and $\delta_{\textrm{b}} = n_{\textrm{b}0}/n_{\textrm{i}0}$ are the ratio of equilibrium electron to ion number densities and beam to ion number densities, respectively and $\lambda_{\textrm{De}} = \sqrt{\epsilon_{0}T_{\textrm{e}}/n_{\textrm{i}0}e^{2}}$ is the Debye length. From Eq. (\ref{eqn_screening_length}), it is evident that the contribution of positive ion beam in dusty plasmas reduces the dust screening length $\lambda_{\textrm{D}}$.
\par 
Typically, positive ions are lost from the ion fluid due to their absorption by the charged dust grains during the dust charging process. Also, the rapid electron impact ionization of neutral gas produces new ions. Thus, the normalized ion loss rate, $G_{\textrm{l}}\equiv {n_{\textrm{d}0}I_{\textrm{Ti}}}/{en_{\textrm{i}0}\omega_{\textrm{pi}}}$ is obtained as 
\begin{equation}
G_{\textrm{l}} = {\nu_{\textrm{l}}N_{\textrm{i}}}\left[1-\frac{1}{\Gamma}\left(\frac{Z}{\sigma_{\textrm{i}}}\Delta Q-0.4\frac{Z^{2}\lambda_{\textrm{D}}}{\sigma_{\textrm{i}}^{2}\lambda_{\textrm{in}}}\Delta Q^{2}+0.8\frac{Z^{2}\lambda_{\textrm{D}}}{\sigma_{\textrm{i}}^{2}\lambda_{\textrm{in}}}\Delta Q\right)\right]
\label{eqn_normalized_ion_loss}
\end{equation}
with the normalized ion-loss frequency, $\nu_{\textrm{l}} \equiv n_{\textrm{d}0}I_{\textrm{Ti}0}/en_{\textrm{i}0}\omega_{\textrm{pi}}$, given by,
\begin{equation}
\nu_{\textrm{l}} = \frac{r_{\textrm{d}}}{\sqrt{\pi}}\frac{\omega_{\textrm{pi}}\sigma_{\textrm{i}}\Gamma}{Z\text{v}_{\textrm{ti}}}{\left(1+\delta_{\textrm{b}}-\delta_{\textrm{e}}\right)}.
\label{eqn_normalized_ion_loss_rate}
\end{equation}
Here, $N_{\textrm{i}} = n_{\textrm{i}}/n_{\textrm{i}0}$ is the normalized ion density, $\text{v}_{\textrm{ti}} = \sqrt{2T_{\textrm{i}}/m_{\textrm{i}}}$ is the thermal velocity of positive ions, $\omega_{\textrm{pi}} = \sqrt{n_{\textrm{i}0}e^2/\epsilon_{0}m_{\mathrm{i}}}$ is the ion plasma oscillation frequency, and
\begin{equation}
\Gamma = 1+Z/\sigma_{\textrm{i}}+0.4Z^{2}\lambda_{\textrm{D}}/\sigma_{\textrm{i}}^{2}\lambda_{\textrm{in}}. 
\end{equation}
\par 
The basic set of fluid equations describing the dynamics of DIAWs in a positive ion-beam-driven dusty plasma are the following normalized continuity, momentum, and Poisson's equations.
\begin{equation}
\frac{\partial N_{\textrm{i}}}{\partial t}+\frac{\partial\left(N_{\textrm{i}}\text{u}_{\textrm{i}x}\right)}{\partial x} = \nu_{\textrm{i}}N_{\textrm{e}} - G_{\textrm{l}},
\label{eqn_continuity}
\end{equation}
\begin{equation}
\begin{aligned}
 \frac{\partial \text{u}_{\textrm{i}x}}{\partial t} + \text{u}_{\textrm{i}x}\frac{\partial \text{u}_{\textrm{i}x}}{\partial x} = -\frac{\partial \psi}{\partial x} + \omega_{\textrm{i}}\text{u}_{\textrm{i}y}{\text{sin}\theta}-\text{u}_{\textrm{i}x}k_{\textrm{n}}-\frac{\sigma_{\textrm{i}}}{N_{\textrm{i}}}\frac{\partial N_{\textrm{i}}}{\partial x}\\-\frac{\text{u}_{\textrm{i}x}}{N_{\textrm{i}}}\left(\nu_{\textrm{i}}N_{\textrm{e}}-G_{\textrm{l}}\right),
 \label{eqn_momentum_xcomponent}
 \end{aligned}
\end{equation}
\begin{equation}
\begin{aligned}
\frac{\partial \text{u}_{\textrm{i}y}}{\partial t} + \text{u}_{\textrm{i}x}\frac{\partial \text{u}_{\textrm{i}y}}{\partial x} =  \omega_{\textrm{i}}\text{u}_{\textrm{i}z}{\text{cos}\theta}-\omega_{\textrm{i}}\text{u}_{\textrm{i}x}{\text{sin}\theta}-\text{u}_{\textrm{i}y}k_{\textrm{n}}\\-\frac{\text{u}_{\textrm{i}y}}{N_{\textrm{i}}}\left(\nu_{\textrm{i}}N_{\textrm{e}}-G_{\textrm{l}}\right),
\label{eqn_momentum_ycomponent}  
\end{aligned}
\end{equation}
\begin{equation}
\frac{\partial \text{u}_{\textrm{i}z}}{\partial t} + \text{u}_{\textrm{i}x}\frac{\partial \text{u}_{\textrm{i}z}}{\partial x} = - \omega_{\textrm{i}}\text{u}_{\textrm{i}y}{\text{cos}\theta}-\text{u}_{\textrm{i}z}k_{\textrm{n}}-\frac{\text{u}_{\textrm{i}z}}{N_{\textrm{i}}}\left(\nu_{\textrm{i}}N_{\textrm{e}}-G_{\textrm{l}}\right),
\label{eqn_momentum_zcomponent}  
\end{equation}
\begin{equation}
\frac{\partial^{2} \psi}{\partial x^{2}} = N_{\textrm{e}}-N_{\textrm{i}}-N_{\textrm{b}}-Z_{\textrm{d}0}\delta_{\textrm{d}}\left(\Delta Q-1\right).
\label{eqn_normalized_Poisson's}
\end{equation}

We assume the electrons to be in thermal equilibrium and they obey the Maxwellian distribution, whereas the streaming positive ion beams follow the drifting Maxwellian distribution (with beam velocity greater than thermal velocity of ion beam). Thus, the electron and ion beam number densities are
\begin{equation}
 N_{\textrm{e}} = \delta_{\textrm{e}}\hspace{0.1cm}\text{exp}(\psi),
\label{eqn_normalized_electron_density}
\end{equation}
\begin{equation}
 N_{\textrm{b}} = \delta_{\textrm{b}}\hspace{0.1cm}\text{exp}(-\psi/\sigma_{\textrm{b}}).
\label{eqn_normalized_electron_density}
\end{equation}
In Eqs. \eqref{eqn_continuity}-\eqref{eqn_momentum_zcomponent},  $\nu_{\textrm{i}}$ is the electron impact ionization frequency normalized by the ion plasma frequency, $(u_{\textrm{i}x},~u_{\textrm{i}y},~u_{\textrm{i}z})$ are the components of the ion fluid velocity normalized by the ion-acoustic speed along  the axes, $\omega_{\textrm{i}} = eB/m_{\textrm{i}}\omega_{\textrm{pi}}$ is the normalized ion-cyclotron frequency, $k_{\textrm{n}} \hspace{0.1cm}(= k_{\textrm{in}} + k_{\textrm{id}})$ is the sum of normalized ion-neutral and ion-dust collision frequencies, and $N_{\textrm{e}}$, and $N_{\textrm{b}}$ are, respectively, the electron and beam number densities normalized by $n_{\textrm{i}0}$. 
\par 
The dimensionless form of the dust charging equation (\ref{eqn_dust_charge}) reads
\begin{equation}
\frac{\omega_{\textrm{pi}}}{\nu_{\textrm{ch}}}\frac{d\Delta Q}{dt} = \frac {1}{\nu_{\textrm{ch}}Z_{\textrm{d0}}e}\left(I_{\textrm{Ti}}+I_{\textrm{b}}+I_{\textrm{e}}\right),
\label{eqn_normalized_dust_charge} 
\end{equation}
or, we have
\begin{equation}
\frac{\omega_{\textrm{pd}}}{\nu_{\textrm{ch}}\sqrt{\mu_{\textrm{d}}(1-\delta_{\textrm{e}})}}\frac{d\Delta Q}{dt} = \frac {1}{\nu_{\textrm{ch}}Z_{\textrm{d0}}e}\left(I_{\textrm{Ti}}+I_{\textrm{b}}+I_{\textrm{e}}\right).
\label{eqn_normalized_dust_charge_adiabatic}
\end{equation}
Here, $\mu_{\textrm{d}} = Z_{\textrm{d}0}m_{\textrm{i}}/m_{\textrm{d}}$  and the dust charging frequency $\nu_{\mathrm{ch}}$ is given by \cite{vladimirov1994scattering}
\begin{equation}
\nu_{\textrm{ch}} = -\frac{1}{Z_{\textrm{d}0}e}\left[\frac{\partial \left(I_{\textrm{Ti}}+I_{\textrm{b}}+I_{\textrm{e}}\right)}{\partial \Delta Q}\right]_{\psi \to 0,\hspace{0.1cm} \Delta Q \to 0},
\label{eqn_dust_charge_frequency}
\end{equation}
which yields
\begin{equation}
\nu_{\textrm{ch}} = \frac{r_{\textrm{d}}}{\sqrt{\pi}}\frac{\omega_{\textrm{pi}}^{2}}{\text{v}_{\textrm{ti}}}\chi_1,
\label{eqn_dust_chrging}
\end{equation}
where $\chi_1$ is given by
\begin{equation*}
\chi_1 = 1+0.8\frac{Z\lambda_{\textrm{D}}}{\sigma_{\textrm{i}}\lambda_{\textrm{in}}}+\sigma_{\textrm{i}}\Gamma+\delta_{\textrm{b}}\sqrt{\frac{\pi\sigma_{\textrm{i}}}{8}}\left[(1+Z)\frac{2m_\textrm{i}}{m_{\textrm{b}}u_{\textrm{b}0}}+u_{\textrm{b}0}\right].
\end{equation*}
\section{NONLINEAR EVOLUTION EQUATIONS} \label{sec-nonl}
To study the nonlinear propagation of  DIAWs in collisional magnetized  ion-beam driven dusty plasmas  we employ the standard reductive perturbation technique (RPT) in which the independent variables are stretched as \cite{ghosh2001small}
\begin{equation}
\xi = \epsilon^{1/2}\left(\text{x}-
\text{v}_{0}\hspace{0.03cm}t \right),\hspace{0.1cm} \tau = \epsilon^{3/2}t,
\label{eqn_stretched_coordinates}
\end{equation}
where $\text{v}_{0}$ is the phase velocity of DIAWs normalized by $C_{\textrm{s}}$, and  $\epsilon~(<1)$ is a positive scaling parameter, which measures the weakness of wave nonlinearity. The dependent variables are expanded in power series of $\epsilon$ as \cite{ghosh2001small}
\begin{eqnarray}
\begin{aligned}
N_{\textrm{i}} = 1 +\epsilon N_{\textrm{i}}^{(1)} +\epsilon^{2}N_{\textrm{i}}^{(2)}+\cdots,\\
u_{\textrm{i}x,\textrm{i}z} = \epsilon u_{\textrm{i}x,\textrm{i}z}^{(1)}+\epsilon^{2}u_{\textrm{i}x,\textrm{i}z}^{(2)}+\cdots,\\
\text{u}_{\textrm{i}y} = \epsilon^{3/2} \text{u}_{\textrm{i}y}^{(1)} + \epsilon^{5/2}\text{u}_{\textrm{i}y}^{(2)}+\cdots,\\
\psi = \epsilon \psi^{(1)}+\epsilon^{2}\psi^{(2)}+\cdots,\\
\Delta Q = \epsilon \Delta Q^{(1)} +\epsilon^{2} \Delta Q^{(2)} +\cdots.
\label{eqn_dependent_variables}
\end{aligned}   
\end{eqnarray}

In the expansion of Eq. (\ref{eqn_dependent_variables}), we have considered the ion gyro-motion as a higher-order effect than the linear. So, we have expanded the velocity perturbation transverse to the magnetic field ($u_{iy}$) with a higher-order of $\epsilon$ than the components parallel to the magnetic field. It follows that the magnetic field will not contribute to the linear dispersion relation for the wave phase velocity. Furthermore, based on observed values \cite{wang2001ionization}, the collision frequency and the frequencies of ion creation and ion loss can be scaled as
\begin{equation}
\begin{aligned}
k_{\textrm{n}} = \epsilon^{3/2}g_{\textrm{n}},~\nu_{\textrm{i}} = \epsilon^{3/2}g_{\textrm{i}},~
\nu_{\textrm{l}} = \epsilon^{3/2}g_{\textrm{l}},
\label{eqn_collision_expand}
\end{aligned}
\end{equation}
From Eq. \eqref{eqn_collision_expand}, it is to be noted that like the transverse velocity perturbation, the collisional as well as the ion creation and ion loss frequencies appear in the order of $\epsilon^{3/2}$, i.e., higher than the linear order $(\epsilon)$ because we have assumed the collisional, as well as the ion creation and ion loss effects to contribute to nonlinear DIA waves.  It implies that these dissipation effects will not cause wave damping in the linear regime but will influence the nonlinear DIA solitary waves.  
\par 
Next, we substitute Eqs. (\ref{eqn_stretched_coordinates})-(\ref{eqn_collision_expand}) into Eqs. (\ref{eqn_continuity})-(\ref{eqn_normalized_Poisson's}), and equate from the resulting equations the coefficients of different power of $\epsilon$. From the lowest order of $\epsilon$, we obtain
\begin{equation}
N_{\textrm{i}}^{(1)} = T\psi^{(1)},
\label{eqn_ion_density_first}  
\end{equation}
\begin{equation}
\text{u}_{\textrm{i}x}^{(1)} = T\text{v}_{0}\psi^{(1)},
\label{eqn_xcomp_velocity_first}  
\end{equation}
\begin{equation}
\text{u}_{\textrm{i}y}^{(1)} = \frac{T\text{v}_{0}^{2}\hspace{0.1cm}\tan\theta \sec\theta}{\omega_{\textrm{i}}}\frac{\partial \psi^{(1)}}{\partial \xi},
\label{eqn_ycomp_velocity_first}  
\end{equation}
\begin{equation}
\text{u}_{\textrm{i}z}^{(1)} = {T\text{v}_{0}}\hspace{0.1cm}\tan\theta\psi^{(1)},
\label{eqn_zcomp_velocity_first}  
\end{equation}
\begin{equation}
0 = \delta_{\textrm{e}}\psi^{(1)}-N_{\textrm{i}}^{(1)}+\frac{\delta_{\textrm{b}}}{\sigma_{\textrm{b}}}\psi^{(1)}-Z_{\textrm{d}0}\delta_{\textrm{d}}\Delta Q^{(1)},
\label{eqn_poisson_first_order}
\end{equation}
where $T = 1 / \text{v}_{0}^{2}\sec^{2}\theta - \sigma_{\textrm{i}}$. From Eqs. (\ref{eqn_xcomp_velocity_first})-(\ref{eqn_zcomp_velocity_first}), it is evident that only the transverse velocity component is associated with the ambient magnetic field. From the next order of $\epsilon$, i.e., $\epsilon^{2}$, we get
\begin{equation}
\begin{aligned}
-\text{v}_{0}\frac{\partial N_{\textrm{i}}^{(2)}}{\partial \xi}+\frac{\partial N_{\textrm{i}}^{(1)}}{\partial \tau}+\frac{\partial (\text{u}_{\textrm{i}x}^{(1)}N_{\textrm{i}}^{(1)})}{\partial\xi}+\frac{\partial \text{u}_{\textrm{i}x}^{(2)}}{\partial \xi}=\\ g_{\textrm{i}}\delta_{\textrm{e}}\psi^{(1)}-g_{\textrm{l}}N_{\textrm{i}}^{(1)}+\frac{g_{\textrm{l}}}{\Gamma}\left(\frac{Z}{\sigma_{\textrm{i}}}+0.8\frac{Z^{2}\lambda_{\textrm{D}}}{\sigma_{\textrm{i}}^{2}\lambda_{\textrm{in}}}\right)\Delta Q^{(1)},
\label{eqn_continuity_second}
\end{aligned}
\end{equation}
\begin{equation}
\begin{aligned}
-\text{v}_{0}\frac{\partial \text{u}_{\textrm{i}x}^{(2)}}{\partial \xi}+\frac{\partial \text{u}_{\textrm{i}x}^{(1)}}{\partial \tau}-\text{v}_{0}N_{\textrm{i}}^{(1)}\frac{\partial \text{u}_{\textrm{i}x}^{(1)}}{\partial \xi}+\text{u}_{\textrm{i}x}^{(1)}\frac{\partial \text{u}_{\textrm{i}x}^{(1)}}{\partial\xi}=\\-\frac{\partial \psi^{(2)}}{\partial \xi}-N_{\textrm{i}}^{(1)}\frac{\partial \psi^{(1)}}{\partial \xi}+ \omega_{\textrm{i}}\text{sin}\theta \text{u}_{\textrm{i}y}^{(2)}-\sigma_{\textrm{i}}\frac{\partial N_{\textrm{i}}^{(2)}}{\partial \xi}\\
-g_{\textrm{n}}u_{\textrm{i}x}^{(1)}-u_{\textrm{i}x}^{(1)}(g_{\textrm{i}}\delta_{\textrm{e}}-g_{\textrm{l}}),
\label{eqn_xcomp_velocity_second}
\end{aligned}
\end{equation}
\begin{equation}
\begin{aligned}
-\text{v}_{0}\frac{\partial \text{u}_{\textrm{i}y}^{(1)}}{\partial \xi}= \omega_{\textrm{i}}\text{cos}\theta\text{u}_{\textrm{i}z}^{(2)}-\omega_{\textrm{i}}\text{sin}\theta\text{u}_{\textrm{i}x}^{(2)}+\omega_{\textrm{i}}\text{cos}\theta N_{\textrm{i}}^{(1)}\text{u}_{\textrm{i}z}^{(1)}\\ -\omega_{\textrm{i}}\text{sin}\theta N_{\textrm{i}}^{(1)}\text{u}_{\textrm{i}x}^{(1)},  
\label{eqn_ycomp_velocity_second}
\end{aligned}
\end{equation}
\begin{equation}
\begin{aligned}
-\text{v}_{0}\frac{\partial \text{u}_{\textrm{i}z}^{(2)}}{\partial \xi}+\frac{\partial \text{u}_{\textrm{i}z}^{(1)}}{\partial\tau}+\text{u}_{\textrm{i}x}^{(1)}\frac{\partial \text{u}_{\textrm{i}z}^{(1)}}{\partial\xi}-\text{v}_{0}N_{\textrm{i}}^{(1)}\frac{\partial \text{u}_{\textrm{i}z}^{(1)}}{\partial\xi}= \\-\omega_{\textrm{i}}\text{cos}\theta\text{u}_{\textrm{i}y}^{(2)}-  g_{\textrm{n}}\text{u}_{\textrm{i}z}^{(1)}-u_{\textrm{i}z}^{(1)}(g_{\textrm{i}}\delta_{\textrm{e}}-g_{\textrm{l}}),
\label{eqn_zcomp_velocity_second}
\end{aligned}
\end{equation} 
\begin{equation}
\begin{aligned}
\frac{\partial^{2}\psi^{(1)}}{\partial\xi^{2}}=\delta_{\textrm{e}}\psi^{(2)}+\frac{\delta_{\textrm{e}}}{2}\psi^{(1)^{2}}-N_{\textrm{i}}^{(2)}+\frac{\delta_{\textrm{b}}}{\sigma_{\textrm{b}}}\psi^{(2)}\\ -\frac{\delta_{\textrm{b}}}{2\sigma_{\textrm{b}}^{2}}\psi^{(1)^{2}}- Z_{\textrm{d}0}\delta_{\textrm{d}}\Delta Q^{(2)}.
\label{eqn_dust_charge_second}
\end{aligned}
\end{equation}  
From the above sets of coupled equations for the first-and second-order perturbed quantities, it is evident that the dust charge fluctuations can influence the linear fundamental mode, as well as the nonlinear evolution of DIA waves. From Eqs.  \eqref{eqn_normalized_dust_charge} and \eqref{eqn_normalized_dust_charge_adiabatic}, we also note that the relevant dynamics of DIAWs will change according to when the dust charge frequency becomes smaller or larger than the ion or dust plasma oscillation frequency. However, we will not consider an intermediate regime of the dust charge frequency but restrict our discussions to two limiting cases of interest, namely the nonadiabatic and adiabatic dust charge variations as in Secs. \ref{sec-sub-nad} and \ref{sec-sub-ad}). 
\subsection{Damped solitary wave: Effect of nonadiabatic dust charge variation} \label{sec-sub-nad}
We consider the case when the dust charge frequency is much lower than the ion plasma oscillation frequency, i.e.,   $\nu_{\textrm{ch}}\ll\omega_{\textrm{pi}}$, i.e., the time scale of dust charge fluctuation is much longer than that of ion plasma oscillation.  So, we assume
\begin{equation}
\nu_{\textrm{ch}} / \omega_{\textrm{pi}} = \epsilon^{3/2}g_{\textrm{ch}},
\label{eqn_charging_frequency}
\end{equation}
where $g_{\rm{ch}}$ is of the order of unity.
With the assumption \eqref{eqn_charging_frequency}  and Eq. (\ref{eqn_stretched_coordinates}), Eq. (\ref{eqn_normalized_dust_charge}) reduces to
\begin{equation}
-\epsilon^{1/2}\text{v}_{\textrm{0}}\frac{\partial \Delta Q}{\partial \xi} + \epsilon^{3/2} \frac{\partial \Delta Q}{\partial \tau} =\epsilon^{3/2} \frac {g_{\textrm{ch}}}{\nu_{\textrm{ch}}Z_{\textrm{d0}}e}\left(I_{\textrm{Ti}}+I_{\textrm{b}}+I_{\textrm{e}}\right).
\label{eqn_dust_charge_weakly}
\end{equation}
Next, using Eq. (\ref{eqn_dependent_variables}) and equating the coefficients of $\epsilon$ from  Eq. (\ref{eqn_dust_charge_weakly}), we obtain 
\begin{equation}
\Delta Q^{(1)} = 0.
\label{eqn_dust_charge_first}
\end{equation}
From Eq. (\ref{eqn_dust_charge_first}), it follows that in the nonadiabatic dust charge variation, the dust charge fluctuation may be a higher-order effect than the first-order of smallness. So, it will not influence the linear DIA mode.
Eliminating the first-order quantities, from Eqs. (\ref{eqn_ion_density_first}), (\ref{eqn_poisson_first_order}), and (\ref{eqn_dust_charge_first}),   we obtain the following expression for the wave phase velocity.
\begin{equation}
\text{v}_{0} = \cos\theta \sqrt{\sigma_{\textrm{i}}+\frac{\sigma_{\textrm{b}}}{\delta_{\textrm{e}}\sigma_{\textrm{b}}+\delta_{\textrm{b}}}}.
\label{eqn_phase_velocity_weakly}
\end{equation}
Equation (\ref{eqn_phase_velocity_weakly}) shows that the dust charge fluctuation does not influence the phase velocity of DIAWs. However, it is significantly modified by the presence of a positive ion beam and the obliqueness of wave propagation about the magnetic field. Furthermore, while the phase velocity gets reduced by the effect of the obliqueness parameter $\theta$, it is enhanced by the ion beam thermal energy. 
\par 
Next, substituting Eq. (\ref{eqn_dependent_variables}) into Eq. (\ref{eqn_dust_charge_weakly}) and equating the coefficients of $\epsilon^{2}$ from the resulting equation, we obtain 
\begin{equation}
\frac{\partial \Delta Q^{(2)}}{\partial \xi} = -\frac{1}{\text{v}_{0}}\left(\frac{T\sigma_{\textrm{i}}g_{\textrm{ch}}\Gamma }{Z\chi_1}-\chi_{2}\right)\psi^{(1)},
\label{eqn_weakly_dust_charge_second}
\end{equation}
where $\chi_2$ is given by
\begin{equation*}
\chi_2 = \frac{\sigma_{\textrm{i}}g_{\textrm{ch}}}{Z\chi_1}\left[\Gamma+\frac{\delta_{\textrm{b}}u_{\textrm{b}0}}{\sigma_{\textrm{b}}}(1+\sigma_{\textrm{b}})\left(1+\frac{2Zm_{\textrm{i}}}{m_{\textrm{b}}u_{\textrm{b}0}^{2}}\right) \sqrt{\frac{\pi}{8\sigma_{\textrm{i}}}}\right]
\end{equation*}
Eliminating ${\partial N_{\textrm{i}}^{(2)}}/{\partial \xi}$, ${\partial \text{u}_{\textrm{i}x}^{(2)}}/{\partial \xi}$, ${\partial \text{u}_{\textrm{i}y}^{(2)}}/{\partial \xi}$, ${\partial \text{u}_{\textrm{i}z}^{(2)}}/{\partial \xi}$, and ${\partial \Delta Q^{(2)}}/{\partial \xi}$ from Eqs. \eqref{eqn_continuity_second}-\eqref{eqn_dust_charge_second}  and   (\ref{eqn_weakly_dust_charge_second}), we obtain the following damped Korteweg-de Vries (KdV) equation.
\begin{equation}
\frac{\partial \psi^{(1)}}{\partial \tau}+P\psi^{(1)} \frac{\partial \psi^{(1)}}{\partial \xi}+R\frac{\partial^{3} \psi^{(1)}}{\partial \xi^{3}}+S\psi^{(1)}=0,
\label{eqn_kdv_modified}
\end{equation}
where the coefficients $P$, $R$, and $S$, respectively, appear due to the wave nonlinearity, dispersion, and damping (dissipation) effects.
Their expressions are given by   $P = \chi_{5}/\chi_{4}$, $R = \chi_{6}/\chi_{4}$, and $S = \chi_{3}/\chi_{4}$, where $\chi_{3}$, $\chi_{4}$, $\chi_{5}$, and $\chi_{6}$ are  
\begin{equation*}
\chi_{3} = T\text{v}_{0}\hspace{0.1cm}\text{sec}^{2}\theta\left(Tg_{\textrm{n}} + g_{\textrm{i}}\delta_{\textrm{e}}(T-1)\right)-\frac{Z_{\textrm{d}0}\delta_{\textrm{d}}}{\text{v}_{0}}\left(\frac{Tg_{\textrm{ch}}\sigma_{\textrm{i}}\Gamma}{Z\chi_{1}}-\chi_{2}\right),
\end{equation*}
\begin{equation*}
\chi_{4 } = 2\text{v}_{0}T^{2},
\end{equation*}
\begin{equation*}
\chi_{5} = \frac{\delta_{\textrm{b}}}{\sigma_{\textrm{b}}^{2}}-\delta_{\textrm{e}}+T^{3}\left(3\text{v}_{0}^{2}\hspace{0.1cm}\text{sec}^{2}\theta-\sigma_{\textrm{i}}\right),
\end{equation*}
\begin{equation*}
\chi_{6} = 1+\frac{T^{2}\text{v}_{0}^{4}}{\omega_{\textrm{i}}^{2}}\text{sec}^{2}\theta\hspace{0.1cm}\text{tan}^{2}\theta.
\end{equation*}
From the expressions of $P$, $R$, and $S$, we find that these coefficients are significantly modified by the presence of ion beam. Also, as expected, the static magnetic field only contributes to the wave dispersion and it influences inversely to it. The wave damping term proportional to $S$ appears due to ion creation, ion loss, and ion-neutral collisions, as well as the dust charging rate.
\par 
To obtain a traveling wave (soliton) solution of Eq. \eqref{eqn_kdv_modified}, we apply the transformation:
\begin{equation} \label{eq-trans}
 \eta = \xi-U(\tau)\tau\equiv \epsilon^{1/2} \left[x-\left(\rm{v}_0+\epsilon U(\tau)\right)t \right]
\end{equation}
 and use the boundary conditions: $\psi^{(1)} \to 0$, $d\psi^{(1)}/d\eta \to 0$, $d^{2}\psi^{(1)}/d\eta^{2} \to 0$ as $|\eta| \to \infty$. Equation \eqref{eq-trans} shows that the $\eta$-frame of reference corresponds to a small increment of the phase velocity $\rm{v}_0$ of DIAWs in the $\xi$-frame. In particular, in absence of any dissipation ($S = 0$), Eq. (\ref{eqn_kdv_modified}) reduces to the KdV equation whose soliton solution is given by
\begin{equation}
\psi^{(1)}\left(\xi, \tau\right) = \psi^{(1)}_\textrm{SA} \text{sech}^{2}\left(\frac{\xi-U_{0}\tau}{\Delta_{\textrm{SW}}}\right),
\label{eqn_analytical_solitary}
\end{equation}
where the soliton amplitude and width are constants, given by, $\psi^{(1)}_\textrm{SA} = 3U_{0}/P$ and $\Delta_{\textrm{SW}} =(4R/U_{0})^{1/2}$. Also, $U_{0}$ is the constant Mach number, which can be regarded as the value of $U(\tau)$  at an initial time $\tau = \tau_{0}$ (with no damping or dissipation effect). 
\par 
To find a soliton solution of Eq. \eqref{eqn_kdv_modified}, we first note that it conserves the total number of particles. This can be verified by integrating Eq. \eqref{eqn_kdv_modified} with respect to $\xi$ and using the boundary conditions stated before as 
\begin{equation} \label{eq-conser1}
\left(\frac{\partial}{\partial\tau}+S\right)\int^{\infty}_{-\infty}\psi^{(1)}(\xi,\tau)d\xi=0.
\end{equation} 
Next, multiplying Eq. \eqref{eqn_kdv_modified} by $\psi^{(1)}$ and integrating over $\xi$ with the same boundary conditions, we get 
\begin{equation} \label{eq-conser2}
\left(\frac{\partial}{\partial\tau}+2S\right)\int^{\infty}_{-\infty}\left[\psi^{(1)}(\xi,\tau)\right]^2d\xi=-R\int^{\infty}_{-\infty}\left(\frac{d\psi^{(1)}}{d\xi}\right)^2d\xi.
\end{equation} 
Since we have DIAWs with positive dispersion, i.e., $R>0$. So, Eq. \eqref{eq-conser2} shows that the DIA soliton energy decays with time (since the right-hand integral is positive definite), implying that the DIA solitons get damped by the effects of the dissipation due to ion-neutral collision, ion creation and ion loss, and the dust charge fluctuations. Thus, a soliton solution of the damped KdV equation (\ref{eqn_kdv_modified}) with a time-dependent amplitude can be obtained as \cite{acharya2024-pop}
\begin{equation}
\psi^{(1)}\left(\xi, \tau\right)= \psi_{\textrm{SW}}^{(1)} (\tau) \hspace{0.1cm}\text{sech}^{2}\left(\frac{\xi-U(\tau)\tau}{\Delta_{\textrm{SW}}(\tau)}\right), 
\label{eqn_time_dependence_analytical}
\end{equation}
where the time-dependent solitary wave amplitude and width, respectively, are
\begin{equation}
\psi_{\textrm{SW}}^{(1)}(\tau) = \frac{3U(\tau)}{P},
\label{eqn_solitary_amplitude}
\end{equation}
\begin{equation}
\Delta_{\textrm{SW}}(\tau) = \left({\frac{4R}{U(\tau)}}\right)^{1/2}.
\label{eqn_solitary_width}
\end{equation} 
Using Eq. (\ref{eqn_time_dependence_analytical}), the energy integral (\ref{eqn_energy_conserved}), given by,
\begin{equation}
E_{\textrm{g}} = \int_{-\infty}^{\infty}|\psi^{(1)}\left(\xi,\tau\right)|^{2}d\xi
\label{eqn_energy_conserved}
\end{equation}
yields
\begin{equation}
E_{\textrm{g}}  = \frac{4}{3}\left[\psi_{\textrm{SW}}^{(1)}(\tau)\right]^{2}\Delta_{\textrm{SW}}(\tau).
\label{eqn_integral_conserved}
\end{equation}
Equation (\ref{eqn_integral_conserved}) shows that the soliton energy is directly proportional to the squared amplitude and width of the soliton, implying that a small decrement of the wave amplitude will result in a significant amount of energy decay compared to the width. 
Next, using Eqs. \eqref{eq-conser2} (in the case of $S=0$ at $\tau=\tau_0$) and (\ref{eqn_integral_conserved}),  
the time-dependent Mach number can be obtained as:
\begin{equation}
U(\tau) = U_{0}\exp\left[-\frac{4}{3}S\left(\tau-\tau_{0}\right)\right].
\label{eqn_mach_number}
\end{equation}
Equation (\ref{eqn_mach_number}) shows that the soliton Mach number decays exponentially at the order of the damping rate, $S$ with a passage of time, $\tau>\tau_0$. Since the soliton amplitude has direct relation with $U(\tau)$ [Eq. \eqref{eqn_solitary_amplitude}] and the width is inversely to $U(\tau)$ [Eq. \eqref{eqn_solitary_width}], a decrement of $U(\tau)$ leads to a decrease in the soliton amplitude and an increase in the soliton width.  
\subsection{Damped solitary wave: Effect of adiabatic dust charge variation} \label{sec-sub-ad}
In this case, we assume the dust charging frequency is much higher than the dust plasma oscillation frequency  (or the dust charging time is much slower than that of dust plasma oscillation), i.e., $\nu_{\textrm{ch}} \gg \omega_{\textrm{pd}}$. So, the dust charge evolution reaches a steady state and there will be no dissipative effect due to the dust charge variation. Thus, Eq. (\ref{eqn_normalized_dust_charge_adiabatic}) reduces to
\begin{equation}
0 = \frac {1}{\nu_{\textrm{ch}}Z_{\textrm{d0}}e}\left(I_{\textrm{Ti}}+I_{\textrm{b}}+I_{\textrm{e}}\right) 
\label{eqn_equilibrium_current}
\end{equation}
Next, we substitute Eqs. (\ref{eqn_ion_current})-(\ref{eqn_electron_current}) and the expansions (\ref{eqn_stretched_coordinates}) and (\ref{eqn_dependent_variables}) into Eq. (\ref{eqn_equilibrium_current}), and equate the coefficients of $\epsilon$ and $\epsilon^{2}$ from the resulting equation to obtain the following expressions for $\Delta Q^{(1)}$ and $\Delta Q^{(2)}$.
\begin{equation}
\Delta Q^{(1)} = \left(\frac{T\sigma_{\textrm{i}}\Gamma}{Z\chi_1}-\lambda_{1}\right)\psi^{(1)},
\label{eqn_first_dust_charge_adiabatic}
\end{equation}
\begin{equation}
\begin{aligned}
\Delta Q^{(2)} = \frac{\sigma_{\textrm{i}}\Gamma}{Z\chi_1}N_{\textrm{i}}^{(2)}-\lambda_{1}\psi^{(2)}-\lambda_{6}\psi^{(1)^{2}},
\label{eqn_second_dust_charge_adiabatic}
\end{aligned}
\end{equation}
where $\lambda$'s are given by
\begin{equation*}
\lambda_{1} = \frac{\sigma_{\textrm{i}}\Gamma}{Z\chi_{1}} +\sqrt{\frac{\pi\sigma_{\textrm{i}}}{8}} \frac{\delta_{\textrm{b}}u_{\textrm{b}0}(\sigma_{\textrm{b}}+1)}{\sigma_{\textrm{b}}Z\chi_{1}}\left(1+\frac{2m_{\textrm{i}}Z}{m_{\textrm{b}}u_{\textrm{b}0}^{2}} \right),
\end{equation*}
\begin{equation*}
    \lambda_{6} = (T\lambda_{2}+\lambda_{5})\left(\frac{\sigma_{\textrm{i}}T\Gamma}{ Z\chi_{1}}-\lambda_{1}\right)-\lambda_{3}\left(\frac{\sigma_{\textrm{i}}T\Gamma}{Z\chi_{1}}-\lambda_{1}\right)^{2}+\lambda_{4},
\end{equation*}
\begin{equation*}
\lambda_{2} = \frac{1}{\chi_{1}}\left(1+\frac{0.8Z\lambda_{\textrm{D}}}{\sigma_{\textrm{i}}\lambda_{\textrm{in}}}\right),
\end{equation*}
\begin{equation}
\begin{aligned}
\lambda_{3} = \frac{1}{2\chi_{1}}\left[\frac{0.8Z\lambda_{\textrm{D}}}{\sigma_{\textrm{i}}\lambda_{\textrm{in}}}-Z\sigma_{\textrm{i}}\Gamma- \right.\\
\left.\delta_{\textrm{b}}u_{\textrm{b}0}Z\left(\frac{\pi\sigma_{\textrm{i}}}{8}\right)^{1/2}\left(1+\frac{2Zm_{\textrm{i}}}{m_{\textrm{b}}u_{\textrm{b}0}^{2}}\right)\right],
\end{aligned}
\end{equation}
\begin{equation*}
\lambda_{4} = \frac{1}{2\chi_{1}}\left[\frac{\sigma_{\textrm{i}}\Gamma}{Z}+\frac{\delta_{\textrm{b}}u_{\textrm{b}0}(\sigma_{\textrm{b}}^{2}-1)}{Z\sigma_{\textrm{b}}^{2}}\left(\frac{\pi\sigma_{\textrm{i}}}{8}\right)^{1/2}\left(1+\frac{2Zm_{\textrm{i}}}{m_{\textrm{b}}u_{\textrm{b}0}^{2}}\right)\right],
\end{equation*}
\begin{equation*}
\lambda_{5} = \frac{1}{\chi_{1}}\left[\sigma_{\textrm{i}}\Gamma+\delta_{\textrm{b}}u_{\textrm{b}0}\left(\frac{\pi\sigma_{\textrm{i}}}{8}\right)^{1/2}\left(1+\frac{2m_{\textrm{i}}(\sigma_{\textrm{b}}-1)}{m_{\textrm{b}}\sigma_{\textrm{b}}u_{\textrm{b}0}^{2}}\right)\right].
\end{equation*}
From Eqs. (\ref{eqn_ion_density_first}), (\ref{eqn_poisson_first_order}),  and (\ref{eqn_first_dust_charge_adiabatic}), the expression for the phase velocity, $\text{v}_{0}$ of DIAWs in the case of adiabatic dust charge variation can be obtained as 
\begin{equation}
\text{v}_{0} = \cos\theta \sqrt{\sigma_{\textrm{i}}+\frac{\sigma_{\textrm{b}}(Z\chi_{1}+\sigma_{\textrm{i}}Z_{\textrm{d}0}\delta_{\textrm{d}}\Gamma)}{Z\chi_{1}(\delta_{\textrm{e}}\sigma_{\textrm{b}}+\delta_{\textrm{b}}+Z_{\textrm{d}0}\delta_{\textrm{d}}\sigma_{\textrm{b}}\lambda_{1})}}.
\label{eqn_phase_velocity_adiabatic}
\end{equation}
From equation (\ref{eqn_phase_velocity_adiabatic}), it is seen that similar to the case of nonadiabatic dust charge variation [Eq. \eqref{eqn_phase_velocity_weakly}], the phase velocity depends on the obliqueness parameter $\theta$,  the ion to electron temperature ratio $\sigma_{\textrm{i}}$, and the beam to electron temperature ratio $\sigma_{\textrm{b}}$. However, in contrast to Eq. \eqref{eqn_phase_velocity_weakly}, an additional term proportional to $\delta_d$ appears in ${\rm v}_0$ due to the presence of charged dust particles, and it depends on the ion-neutral collision via $\chi_1$. In absence of the charged dust particles (i.e., $\delta_d=0$), Eq. (\ref{eqn_phase_velocity_adiabatic}) becomes identical with Eq. \eqref{eqn_phase_velocity_weakly}. Thus, it follows that in the limit of nonadiabatic dust charge variation, the linear phase velocity of DIAWs in ion-beam-driven dusty plasmas is basically the same as for ion-acoustic waves in ion-beam-driven plasmas without charged dusts. By the same way as for the case of nonadiabatic dust charge variation (See Sec. \ref{sec-sub-nad}), the evolution equation for DIA solitary waves in dusty plasmas with adiabatic dust charge variation can be obtained as 
\begin{equation}
\frac{\partial \psi^{(1)}}{\partial \tau}+P_{1}\psi^{(1)} \frac{\partial \psi^{(1)}}{\partial \xi}+R_{1}\frac{\partial^{3} \psi^{(1)}}{\partial \xi^{3}}+S_{1}\psi^{(1)}=0.
\label{eqn_kdv_modified_adiabatic}
\end{equation}
We note that compared to the case of nonadiabatic dust charge variation, the nonlinear ($P_{1}$), dispersion ($R_{1}$), and damping ($S_{1}$) coefficients are significantly modified, where $P_{1} = \lambda_{8}/\lambda_{7}$, $R_{1} = \lambda_{9}/\lambda_{7}$, and $S_{1}=\lambda_{10}/\lambda_{7}$ with the following expressions for $\lambda_{7}$, $\lambda_{8}$, $\lambda_{9}$, and $\lambda_{10}$:
\begin{equation*}
\lambda_{7} = 2\text{v}_{0}T^{2}\alpha,
\end{equation*}
\begin{equation*}
\lambda_{8} = \alpha T^{3}(3\text{v}_{0}^{2}\hspace{0.1cm}\sec^{2}\theta-\sigma_{\textrm{i}})-\delta_{\textrm{e}}+{\delta_{\textrm{b}}}/{\sigma_{\textrm{b}}^{2}}-2Z_{\textrm{d}0}\delta_{\textrm{d}}\lambda_{6}, 
\end{equation*}
\begin{equation*}
\lambda_{9} =  1+ \frac{\alpha T^{2}\text{v}_{0}^{4}}{\omega_{\textrm{i}}^{2}}  \hspace{0.1cm}\tan^{2}\theta \hspace{0.1cm}\sec^{2}\theta, 
\end{equation*}
\begin{equation*}
\lambda_{10} = \alpha T\text{v}_{0}\sec^{2}\theta \left[Tg_{\textrm{n}}+(T-1)g_{\textrm{i}}\delta_{\textrm{e}}-\beta\right].
\end{equation*}
Here, $\alpha$ and $\beta$ are given by
\begin{equation*}
\alpha = \left(1+Z_{\textrm{d}0}\delta_{\textrm{d}}\sigma_{\textrm{i}}\Gamma/Z\chi_{1}\right),   
\end{equation*}
\begin{equation*}
\beta = \frac{Zg_{\textrm{l}}}{\sigma_{\textrm{i}}\Gamma}\left(1+\frac{0.8Z\lambda_{\textrm{D}}}{\sigma_{\textrm{i}}\lambda_{\textrm{in}}}\right)\left(\frac{\sigma_{\textrm{i}}\Gamma T}{Z\chi_{1}}-\lambda_{1}\right).    
\end{equation*}
\par 
Following the same procedure as for Eq. (\ref{eqn_kdv_modified}), we obtain a traveling wave (soliton) solution of Eq. (\ref{eqn_kdv_modified_adiabatic}) as
\begin{equation}
\psi^{(1)}\left(\xi, \tau\right)= \psi_{\textrm{SA}}^{(1)} (\tau) \hspace{0.1cm}\text{sech}^{2}\left(\frac{\xi-U(\tau)\tau}{\Delta_{\textrm{SA}}(\tau)}\right), 
\label{eqn_time_dependence_analytical_adiabatic}
\end{equation}
where the time-dependent solitary wave amplitude and width, respectively, are 
\begin{equation}
\psi_{\textrm{SA}}^{(1)}(\tau) = \frac{3U(\tau)}{P_{1}},
\label{eqn_solitary_amplitude_adiabatic}
\end{equation}
\begin{equation}
\Delta_{\textrm{SA}}(\tau) = \sqrt{\frac{4R_{1}}{U(\tau)}}.
\label{eqn_solitary_width_adiabatic}
\end{equation} 
 Also, similar to the case of nonadiabatic dust charge variation, the corresponding energy integral and the time-dependent Mach number can be obtained as  
\begin{equation}
E_{\textrm{g}}  = \frac{4}{3}\left[\psi_{\textrm{SA}}^{(1)}(\tau)\right]^{2}\Delta_{\textrm{SA}}(\tau),
\label{eqn_integral_conserved_adiabatic}
\end{equation}
\begin{equation}
U(\tau) = U_0\text{exp}\left(-\frac{4}{3}S_{1}(\tau-\tau_{0}\right).
\label{eqn_mach_number_adiabatic}
\end{equation}
Equations \eqref{eqn_integral_conserved_adiabatic} and \eqref{eqn_mach_number_adiabatic} have the same forms as Eqs. \eqref{eqn_integral_conserved} and \eqref{eqn_mach_number}. So, similar interpretations for the decay of the soliton energy and the Mach number with time $\tau > \tau_{0}$, and for the characteristics of the soliton amplitude and width also apply.
\section{RESULTS AND DISCUSSION} \label{sec-resul}
In this section, we will numerically analyze the characteristics of the dust charge number $(Z_{d0})$, the linear phase velocity $(\text{v}_0)$, the soliton energy $(E_g)$, as well as the profiles of DIA solitons in two different cases, namely nonadiabatic and adiabatic dust charge variations, with different plasma parameter values that are relevant to laboratory plasmas. Specifically, we consider the parameter values as \cite{ghosh2007weakly} \cite{d1997ionization} \cite{nakamura2002experiments} \cite{nakamura2004effects} \cite{misra2011large} \cite{choudhary2020comparative} ion mass, $m_{\textrm{i}}= 40$ amu; mass of streaming positive ion beam, $m_{\textrm{b}} = 4$ amu; ion number density, $n_{\textrm{i}0} = 10^{13}$ m$^{-3}$; number density of positive ion beam, $n_{\textrm{{b}0}}= (0.06-0.60) n_{\textrm{i}0}$; dust number density, $n_{\textrm{d}0}= 10^{10}$ m$^{-3}$; ion temperature, $T_{\textrm{i}}= 0.5$ eV; electron temperature, $T_{\textrm{e}}= 2.5$ eV; positive ion-beam temperature, $T_{\textrm{b}}= 0.01-0.05$ eV; gas temperature$ = 0.0259$ eV; dust-grain radius, $r_{\textrm{d}}= 0.2 ~\mu$m; mass density of dust particle, $\rho_{\textrm{d}}= 2000$ kg/m$^{3}$; magnetic field strength ranges from $0.003$ to $0.014$ T; obliqueness parameter, $\theta= 10^{\circ}-60^{\circ}$;   gas pressure $0.27$ Pa; constant initial Mach number, $U_{0} = 0.01$. To avoid the wave instability or damping due to ion-ion collisions, we have considered the beam velocity, $\text{v}_{\textrm{b}0}~(\geq2C_s)$ to be close to two times the DIA speed or higher \cite{misra2011large}. Unless otherwise mentioned, the fixed parameter values for the magnetic field, obliqueness of wave propagation about the magnetic field, positive ion-beam temperature, beam velocity, and number density of positive ion-beam are considered as $0.014$ T, $30^{\circ}$, $0.05$ eV, $2C_{\textrm{s}}$, and $0.5n_{\textrm{i}0}$  respectively. 
 \par
\begin{figure}[!h]
\centering
\includegraphics[width=8.5cm]{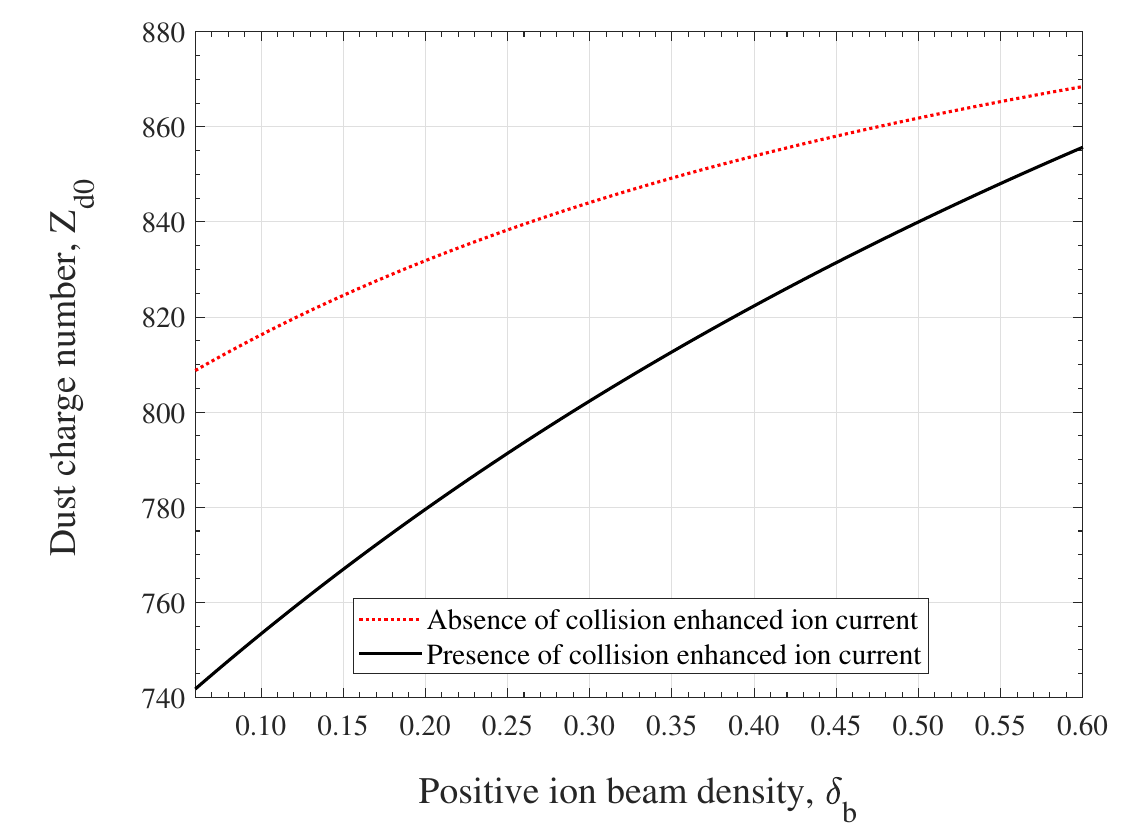}
\caption{Profiles of the dust charge number, $Z_{\textrm{d}0}$ [Eq. (\ref{eqn_dust_charging_equation})] is shown against the ion-beam to ion density ratio, $\delta_{\textrm{b}}$ in two different cases: Absence and presence of collision enhanced ion current.} 
\label{fig_1}
\end{figure}

\begin{figure}[!h]
\centering
\includegraphics[width=8.5cm]{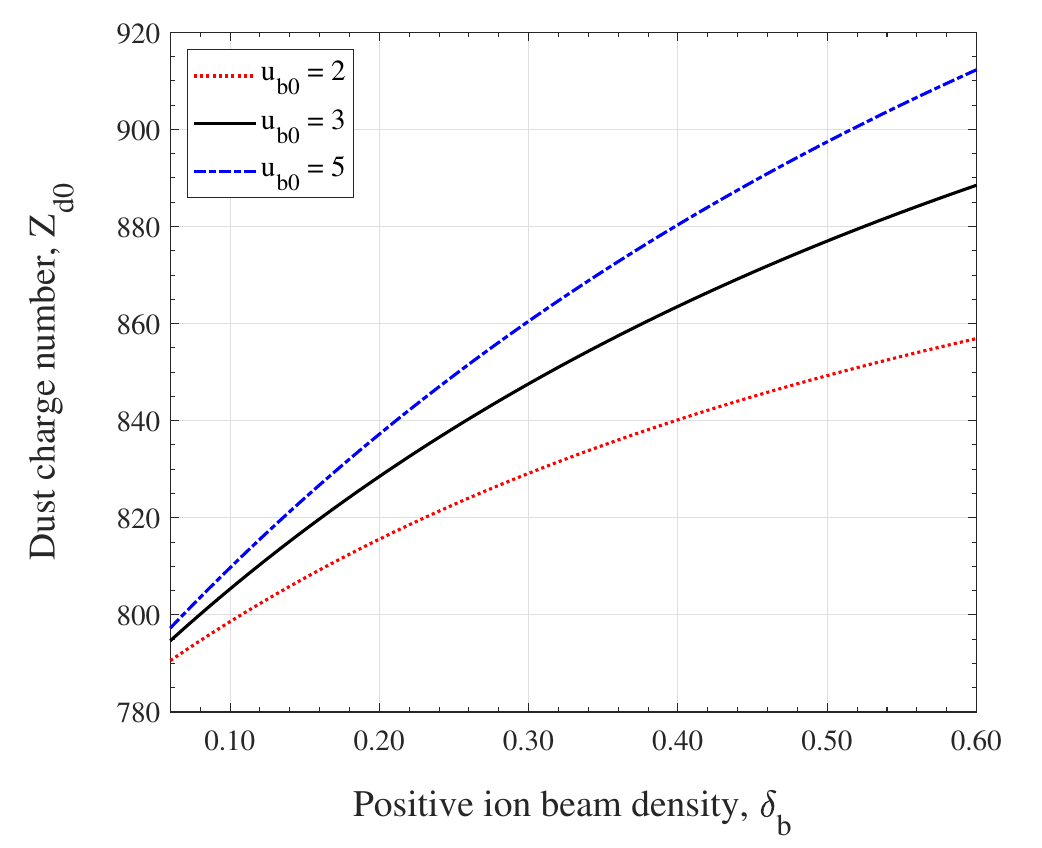}
\caption{Profiles of the dust charge number, $Z_\textrm{{d}0}$ [Eq. (\ref{eqn_dust_charging_equation})] is shown against the ion-beam to ion density ratio, $\delta_{\textrm{b}}$ with different values of the positive ion-beam velocity, $u_{\textrm{b}0}$.}
\label{fig_2}
\end{figure}

 
In what follows, we calculate the dust charge number, $Z_{\textrm{d}0}$ for a given set of parameter values stated above by the Newton-Raphson method and using the following dust charging equation at equilibrium. 
\begin{equation}
I_{\textrm{Ti}0} + I_{\textrm{b}0} + I_{\textrm{e}0} = 0.
\label{eqn_dust_charging_equation}
\end{equation}
The profiles of $Z_{d0}$ against the positive ion-beam to ion density ratio, $\delta_b$ are displayed in Figs. \ref{fig_1} and \ref{fig_2}.  Due to ion-neutral collisions, the number of positive ions residing on the dust-grain surface changes, and consequently, the ion current flowing to the dust grain significantly alters. Figure \ref{fig_1} displays the profiles of $Z_{d0}$ in two cases: in the absence and presence of ion-neutral collision enhanced currents. We have seen that the ion-neutral collision enhanced current significantly reduces the dust charge number (See the solid black line) compared to the case of no ion-neutral collision (See the red-dotted line) for the reason above. Also, in both these cases, the dust charge number tends to increase as the beam density increases. Physically, an increase in the ion-beam density gives rise to an enhancement of the electron concentration to maintain the charge quasineutrality. As a result, the negativity of the dust charge gets enhanced. For example, the dust charge number increases from $809$ to $869$  and $742$ to $855$ in the absence and presence of the collision enhanced ion-current when the beam density increases from $0.06$ to $0.60$, respectively. The obtained dust charge number is consistent with the previous work \cite{el2003dust}, which considered it arbitrary. We also examine the influence of the ion-beam streaming velocity ($u_{b0}$) on the profiles of $Z_{d0}$, as shown in Fig. \ref{fig_2}. The dust charge number increases for the increase in ion beam velocity. Physically, as the streaming velocity of the ion beam increases, the beam flux residing on the dust-grain surface increases. As a result, the electron current increases to maintain the equilibrium of current flows and, hence, increases the dust charge number.
 \par
\begin{figure}[!h]
\centering
\includegraphics[width=8.5cm]{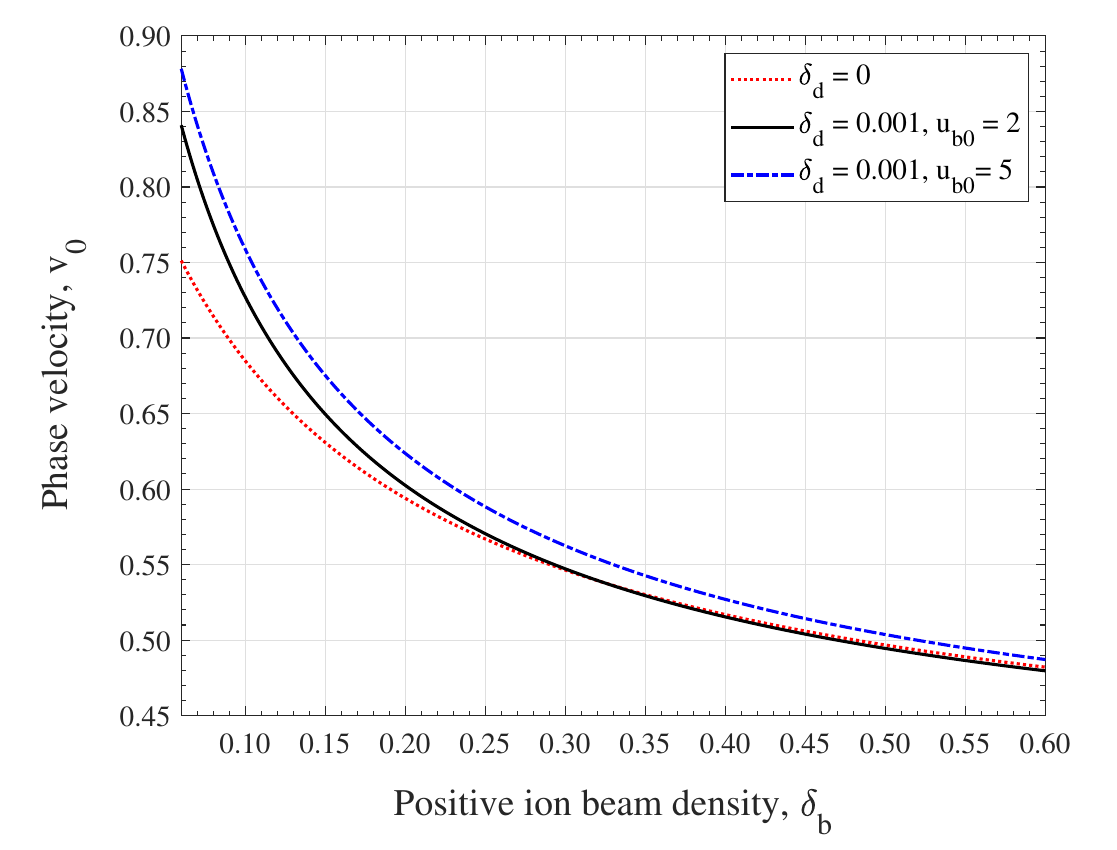}
\caption{Profiles of the phase velocity of DIAWs, $\text{v}_{0}$ [Eq. (\ref{eqn_phase_velocity_adiabatic}), the case of adiabatic dust charge variation] is shown against the ion-beam to ion density ratio, $\delta_{\textrm{b}}$ with different values of the positive ion-beam velocity, $u_{\textrm{b}0}$ and the dust to ion number density ratio, $\delta_{\textrm{d}}$ as in the legends. The value, $\delta_{\textrm{d}} = 0$ corresponds to the phase velocity in the case of nonadiabatic dust charge variation [Eq. \eqref{eqn_phase_velocity_weakly}], which is independent of the ion-beam velocity.}
\label{fig_3}
\end{figure}

 
In what follows, we inspect the characteristics of the phase velocity, $\rm{v}_0$ with the variations of the beam velocity $u_{b0}$ and the dust to ion density ratio $\delta_d$ in the cases of nonadiabatic and adiabatic dust charge variations as displayed in Fig. \ref{fig_3}.  One can recover the profiles of the phase velocity in the nonadiabatic case from the plots of Eq. \eqref{eqn_phase_velocity_adiabatic} in the limit of zero dust density (i.e., $\delta_d=0$). In this case, $\rm{v}_0$ is independent of the dust charge number and the ion-beam velocity. From Fig. \ref{fig_3}, it is observed that the phase velocity tends to slow down as the ion-beam density increases. Also, it is reduced with an enhancement of the beam velocity when we consider the adiabaticity of dust charge variation. In the latter, we also observe an increment of the phase velocity with an enhancement of the dust-number density (See the dotted and solid lines). Thus, it follows that DIAWs in ion-beam-driven dusty plasmas with adiabatic dust charge variation travels faster with longer wavelengths than dusty plasmas with nonadiabatic dust charge variation.  
%%%%%%%%%%%%%%%%%%
%%%%%%%%%%%%%%%%%%%%%%%%%%%%%%%%%%%%%%%%%%%%
\par
\begin{figure}[!h]
\centering
\includegraphics[width=8.5cm]{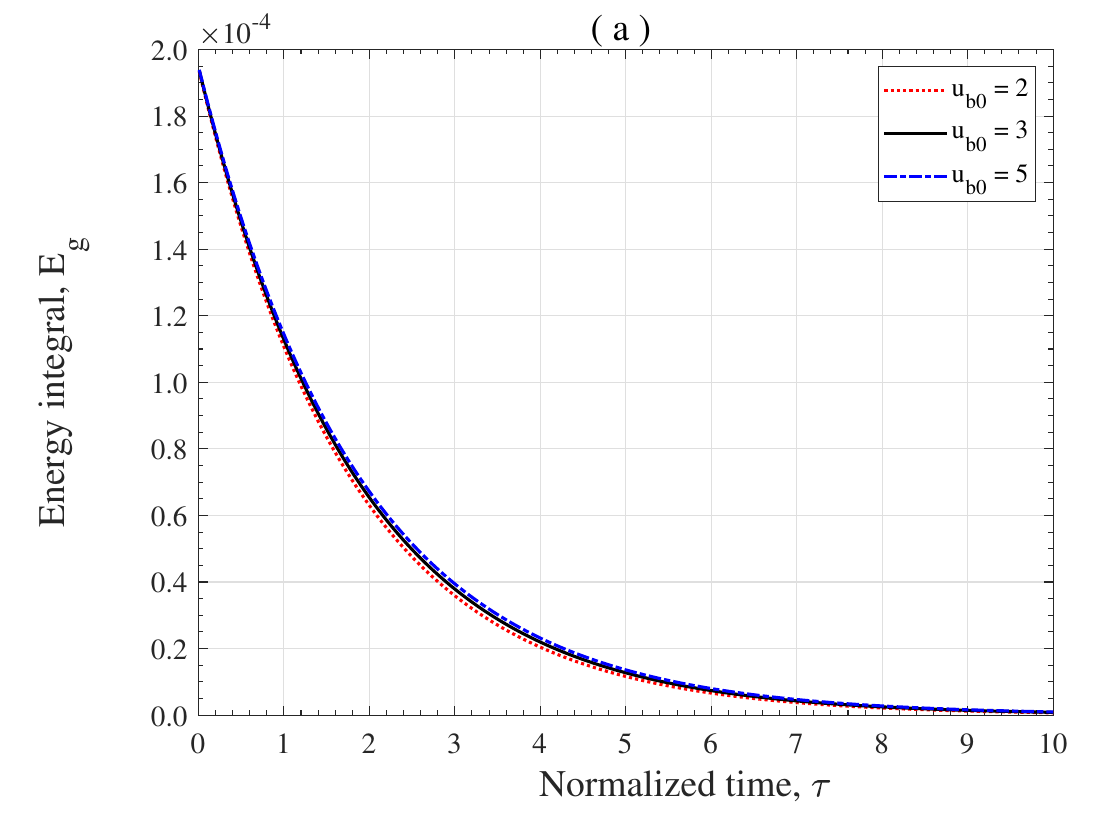}
\quad
\includegraphics[width=8.5cm]{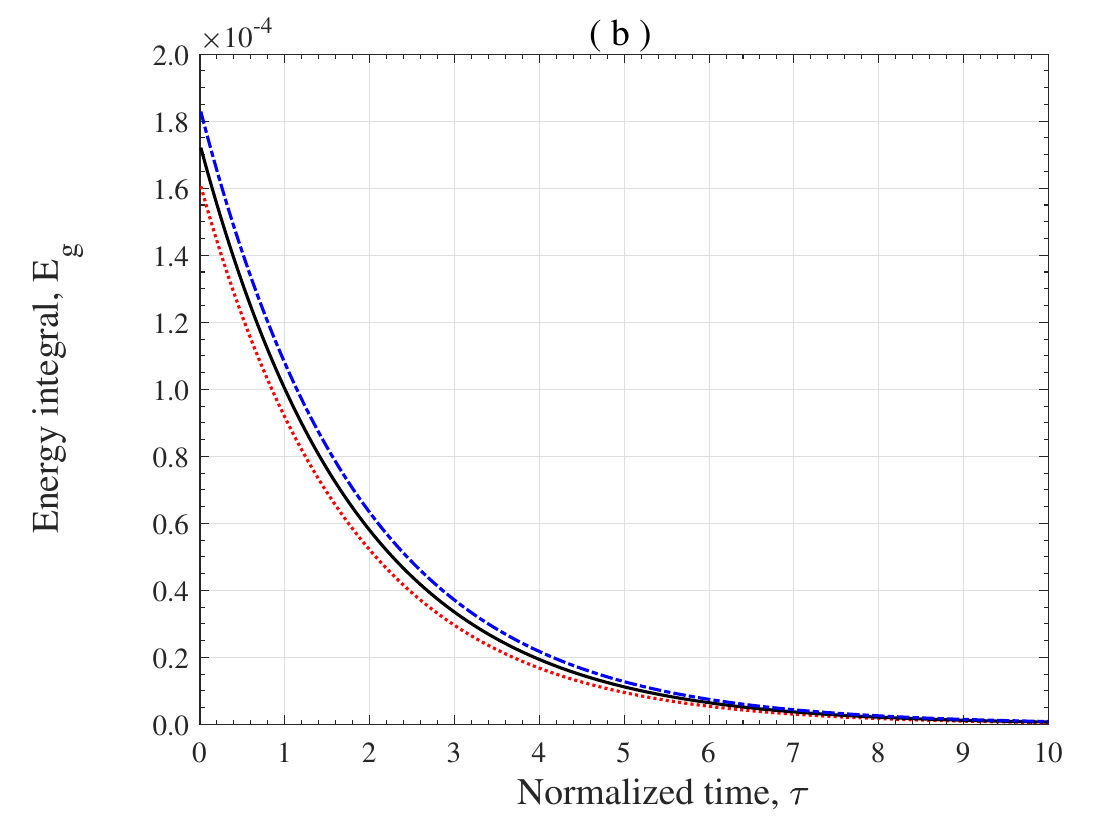}
\caption{Decay of the soliton energy $(E_g)$ with time $\tau$ is shown with different values of the positive ion-beam streaming velocity $u_{b0}$ corresponding to Eqs.  (\ref{eqn_integral_conserved}) [Subplot (a), Case of noadiabatic dust-chare variation] and (\ref{eqn_integral_conserved_adiabatic}) [Subplot (b), Case of adiabatic dust charge variation].}
\label{fig_4}
\end{figure}

 
The variation of the wave energy over time is crucial to diagnosing the growing or decaying nature of soliton amplitudes in plasmas. We have shown the profiles of the soliton energy, $E_g$ in Fig. \ref{fig_4} in the cases of nonadiabatic [Subplot (a)] and adiabatic dust charge variations [Subplot (b)]. We observe that the soliton energy decays with time and becomes zero after a finite time, implying that DIA solitons get damped by the effects of ion creation, ion loss, and ion-neutral collisions and will no longer exist after a finite time. The ion beam has no significant influence on the soliton energy in the case of nonadiabatic dust charge variation. However, when we consider the adiabatic dust charge variation, the soliton energy gets significantly reduced by the effects of the higher beam streaming velocity. Thus, streaming positive ion beam in collisional dusty plasmas induces stronger damping of DIA solitons with dust charge variations.
%%%%%%%%%%%%%%%%%%%%%%%%%%%%%%%%%%%%%% 
\par
\begin{figure}[!h]
\centering
\includegraphics[width=8.5cm]{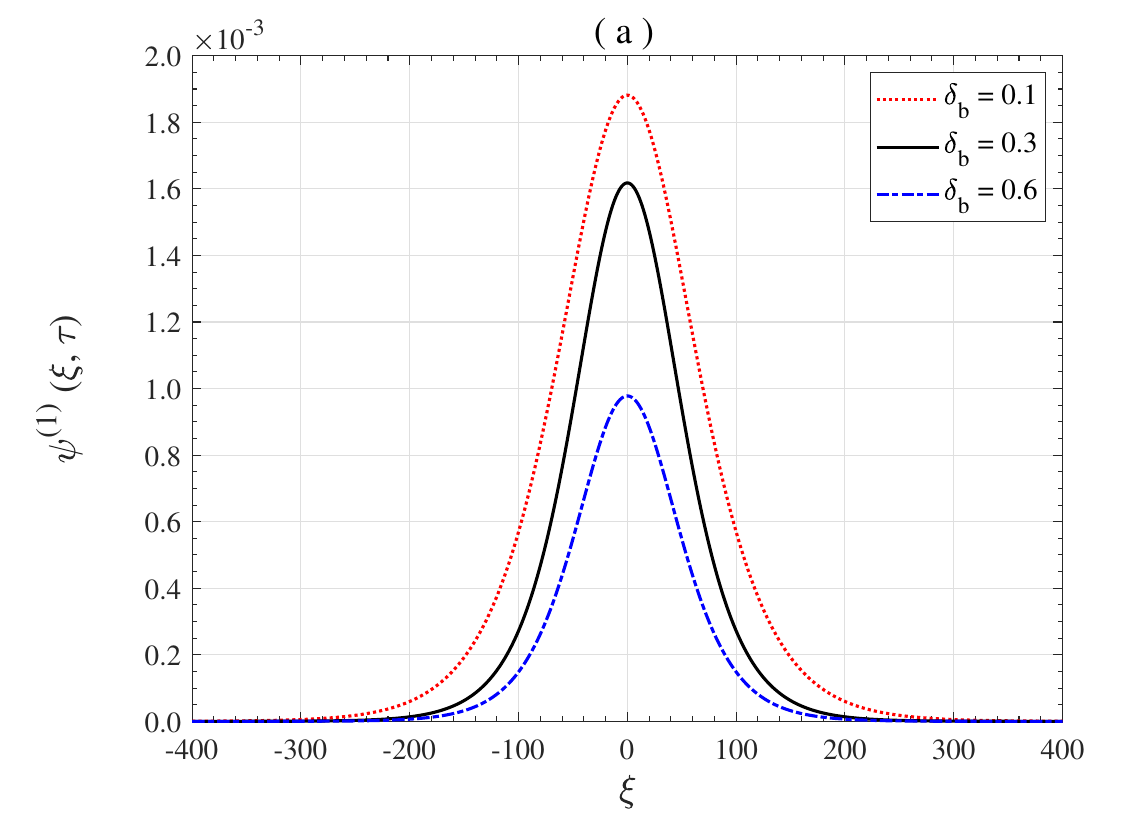}
\quad
%%%%%%%%%%%%%%%%%%%%%%%%%%%%%%%%%%%%%%%%%%%%%%%%%%%%%%%%%%%%%%%%%%%%%%%%%%%%%%%%%%%
\includegraphics[width=8.5cm]{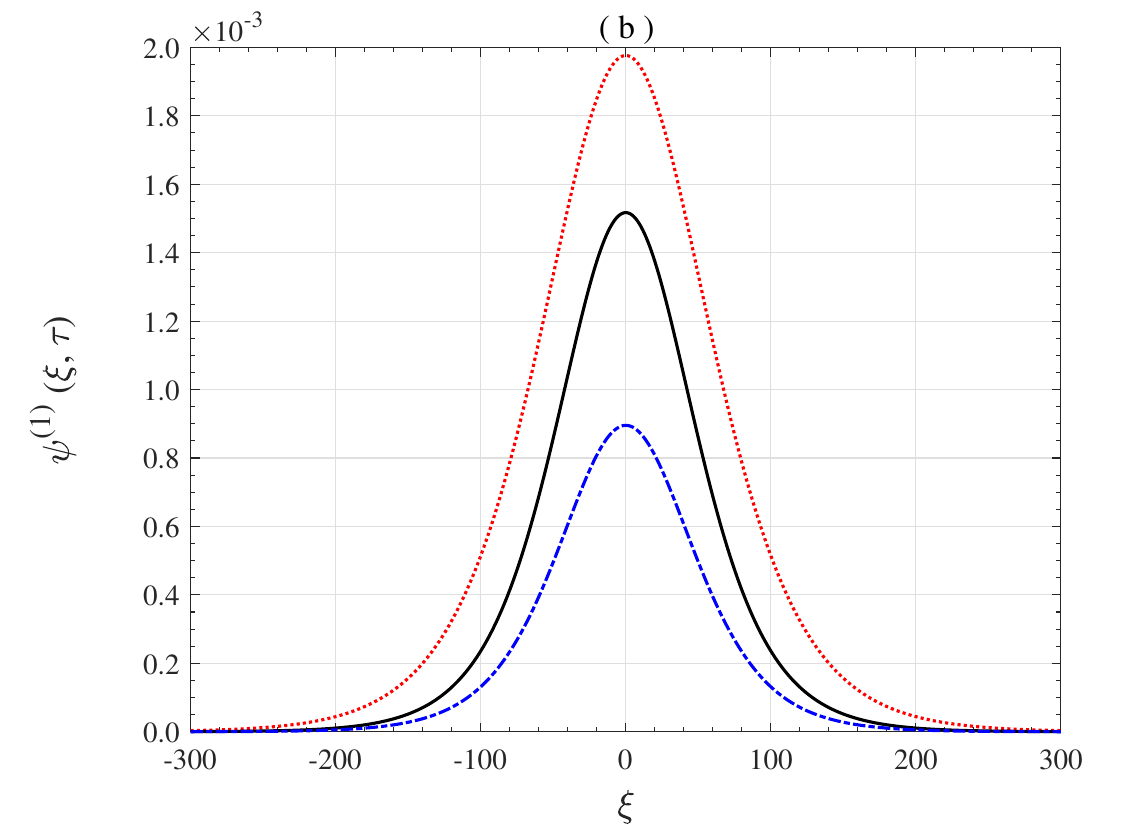}
\caption{Profiles of DIA solitons are shown for different values of the beam to ion density ratio, $\delta_b$. Subplots (a) and (b), respectively, correspond to solutions (\ref{eqn_time_dependence_analytical})  and (\ref{eqn_time_dependence_analytical_adiabatic}) obtained in the cases of nonadiabatic and adiabatic dust charge variations.}
\label{fig_5}
\end{figure}
\begin{figure}[!h]
\centering
\includegraphics[width=8.5cm]{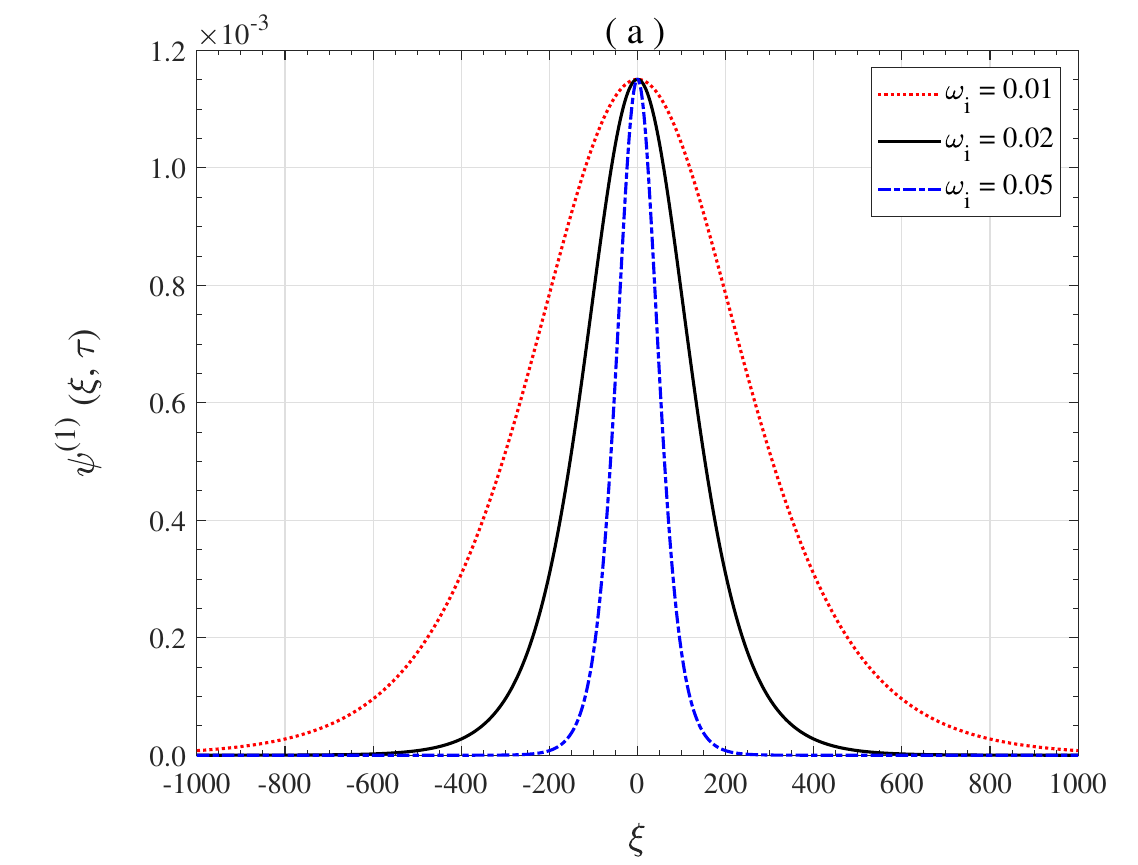}
\quad
\includegraphics[width=8.5cm]{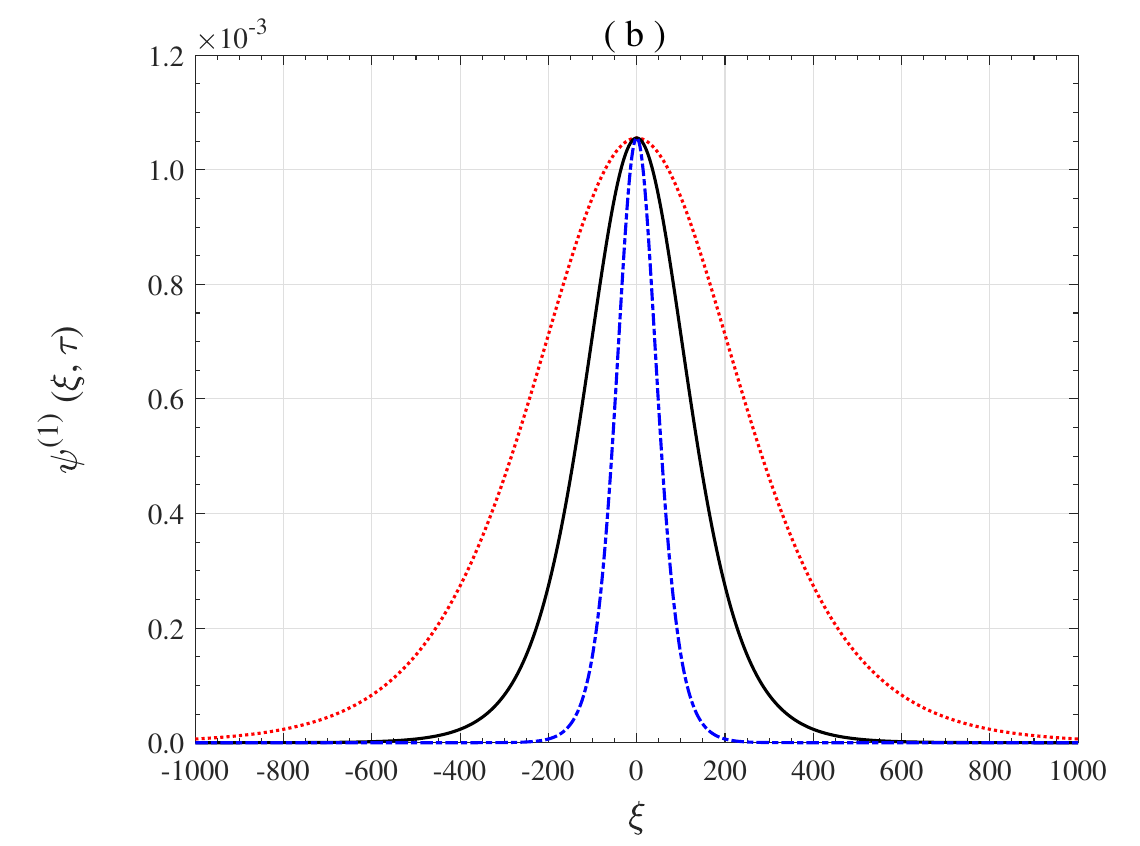}
\caption{Profiles of the DIA solitons are shown for different values of the normalized ion-gyrofrequency $\omega_i$ as in the legends. Subplots (a) and (b) are corresponding to Eqs. \eqref{eqn_time_dependence_analytical} and (\ref{eqn_time_dependence_analytical_adiabatic}) respectively.}
\label{fig_6}
\end{figure}
\begin{figure}[!h]
\centering
\includegraphics[width=8.5cm]{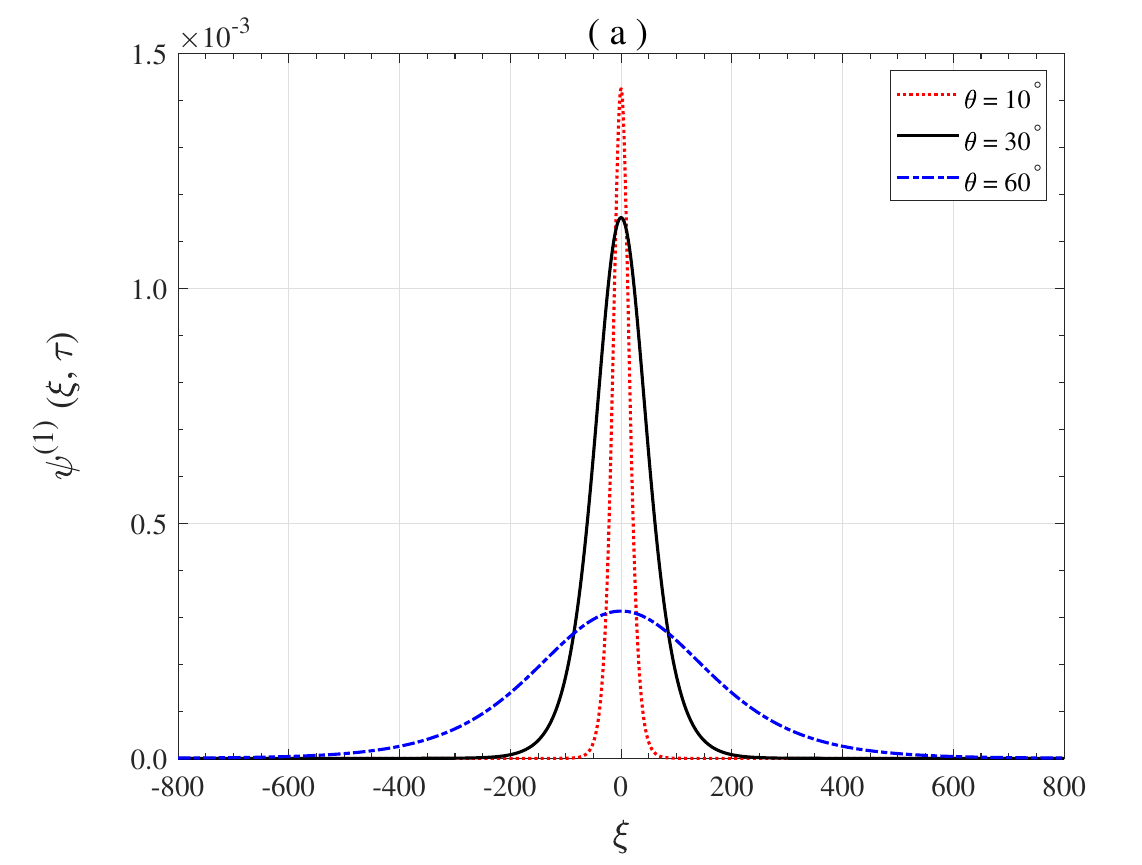}
\quad
\includegraphics[width=8.5cm]{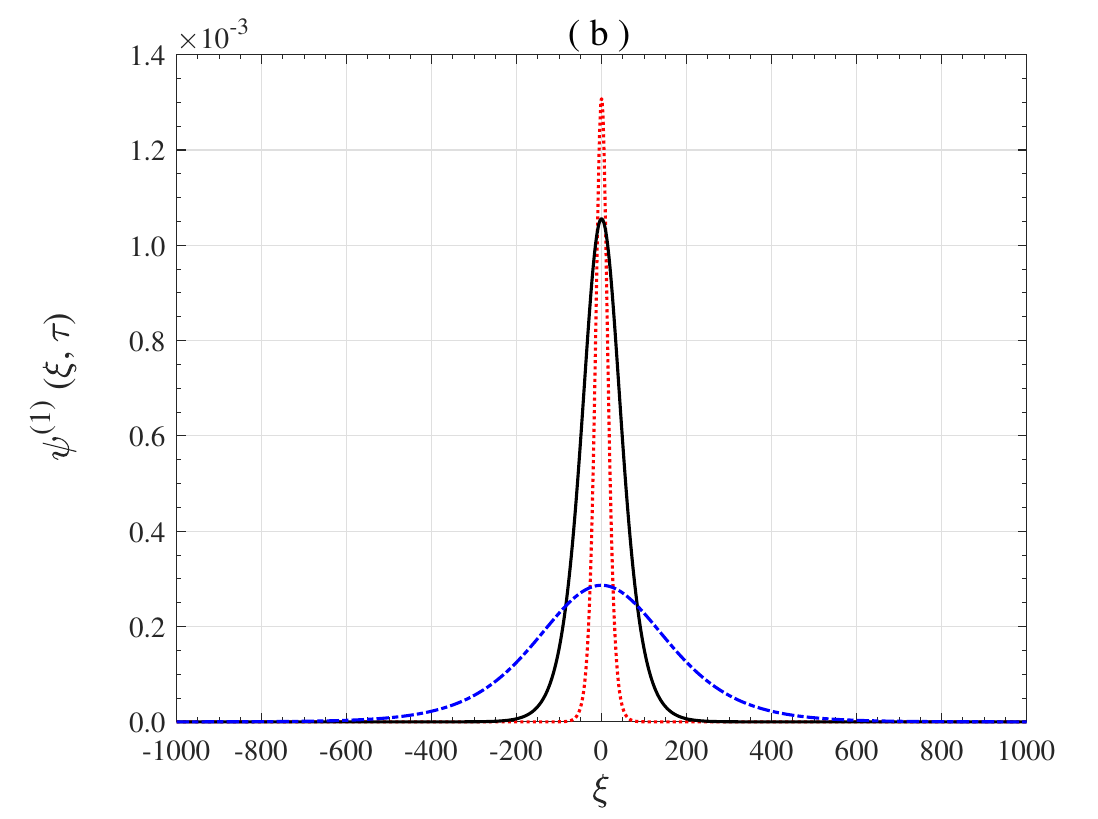}
\caption{Profiles of the DIA solitons are shown for different values of the obliqueness parameter $\theta$ as in the legends. Subplots (a) and (b) are corresponding to Eqs. \eqref{eqn_time_dependence_analytical} and (\ref{eqn_time_dependence_analytical_adiabatic}) respectively.}
\label{fig_7}
\end{figure}
\begin{figure}[!h]
\centering
\includegraphics[width=8.5cm]{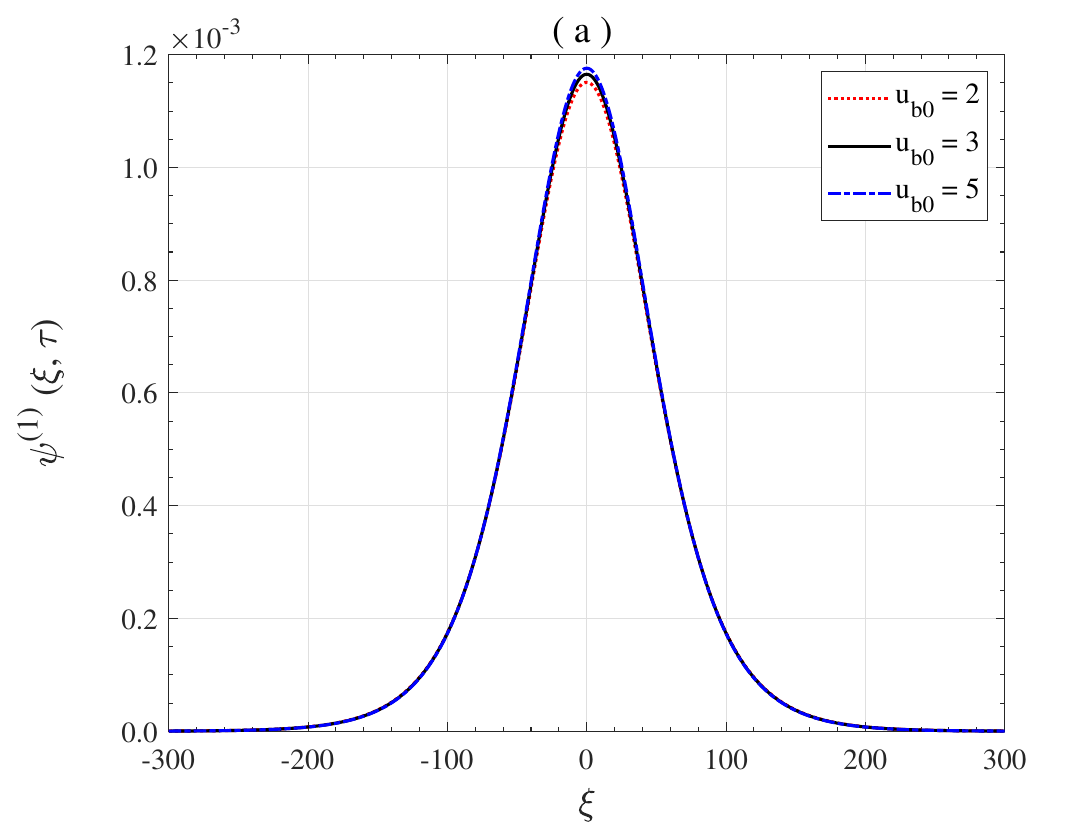}
\quad
\includegraphics[width=8.5cm]{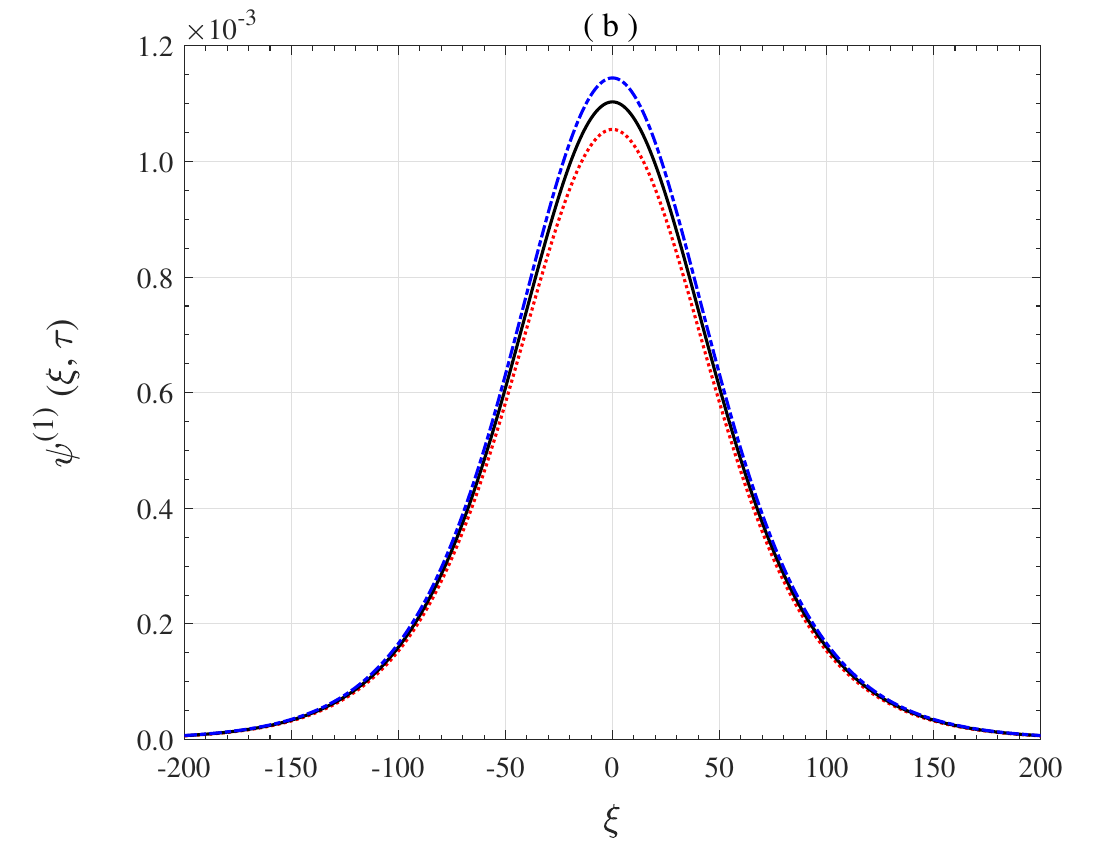}
\caption{Profiles of the DIA solitons are shown for different values of the positive ion-beam velocity $u_{b0}$ as in the legends. Subplots (a) and (b) are corresponding to Eqs. \eqref{eqn_time_dependence_analytical} and (\ref{eqn_time_dependence_analytical_adiabatic}) respectively.}
\label{fig_8}
\end{figure}
\begin{figure}[!h]
\centering
\includegraphics[width=8.5cm]{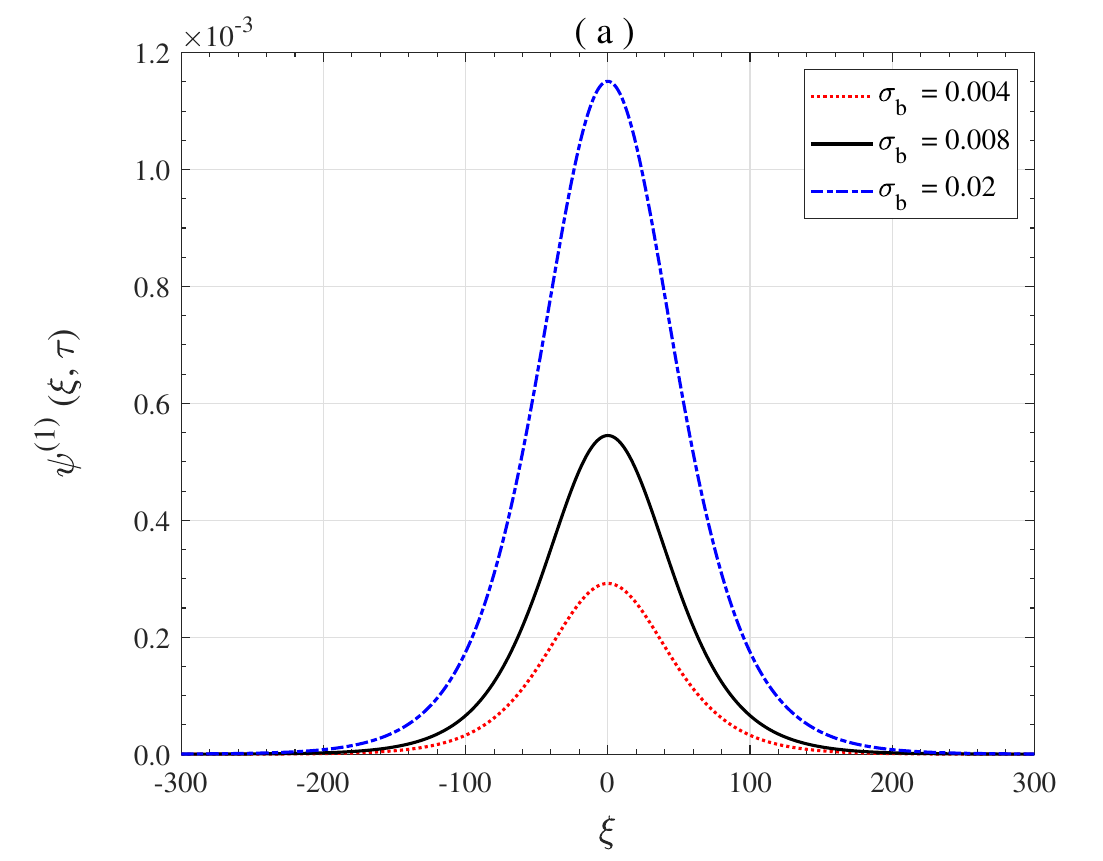}
\quad
\includegraphics[width=8.5cm]{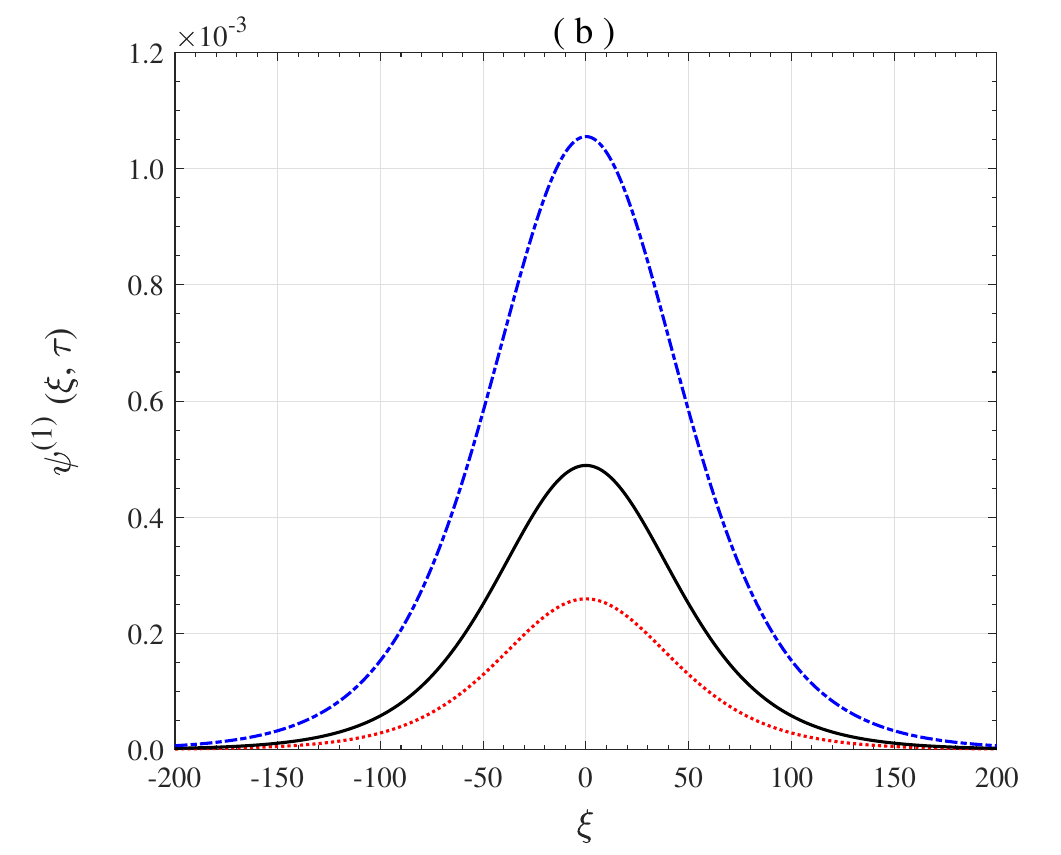}
\caption{Profiles of the DIA solitons are shown for different values of the beam to ion temperature ratio $\sigma_b$ as in the legends. Subplots (a) and (b) are corresponding to Eqs. \eqref{eqn_time_dependence_analytical} and (\ref{eqn_time_dependence_analytical_adiabatic}) respectively.}
\label{fig_9}
%%%%%%%%%%%%%%%%%%%%%%%%%%%%%%%%%
\end{figure}
\begin{figure}[!h]
\centering
\includegraphics[width=8.5cm]{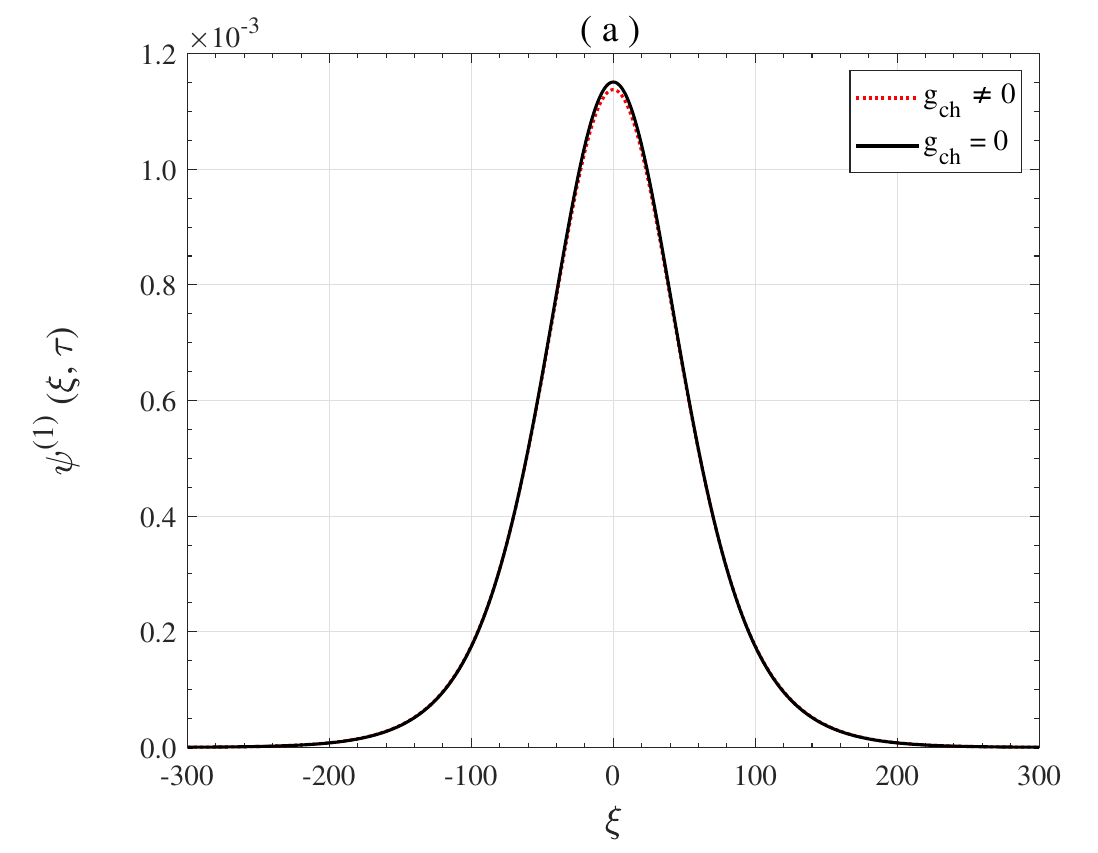}
\quad
\includegraphics[width=8.5cm]{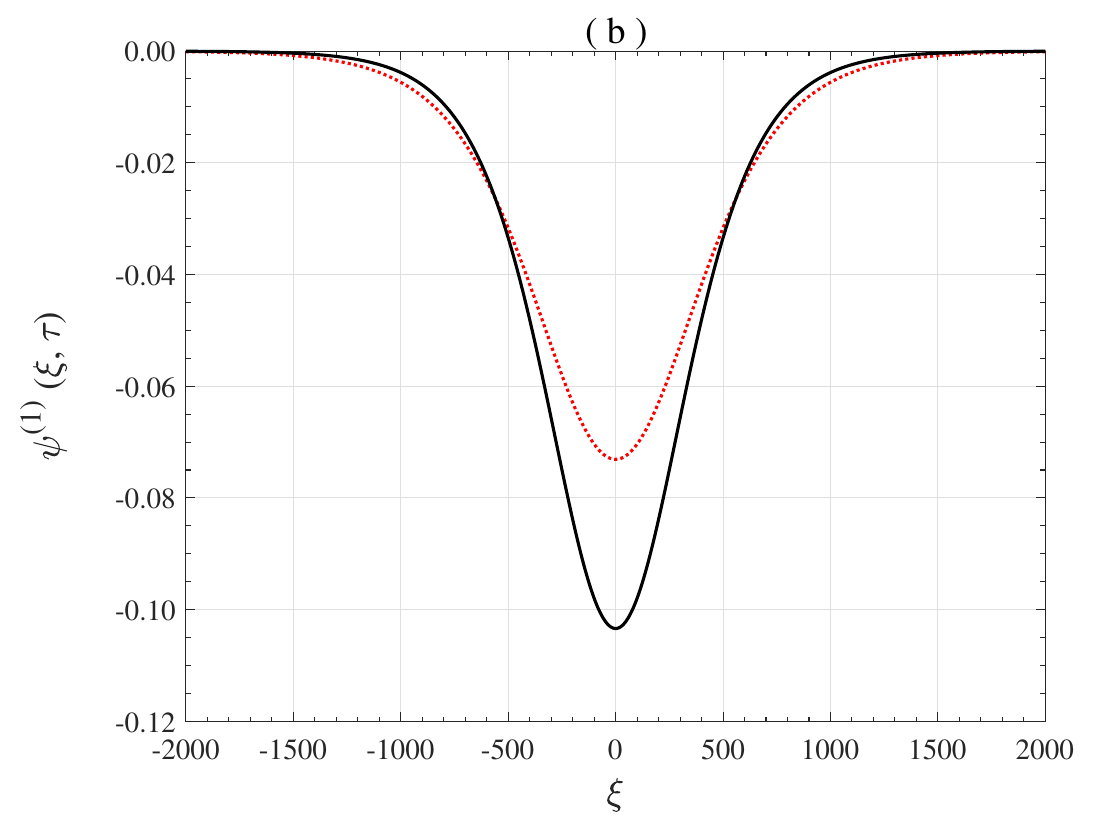}
\caption{Profiles of DIA solitons [Eq. \eqref{eqn_time_dependence_analytical}] are shown in the presence [Subplot (a)] and absence [Subplot (b)] of positive ion-beam density with (i) dust charge fluctuations ($g_{\textrm{ch}} = 0.5$) and (ii) constant dust charge ($g_{\textrm{ch}} = 0$).}
\label{fig_10}
\end{figure}
\begin{figure}[!h]
\centering
\includegraphics[width=8.5cm]{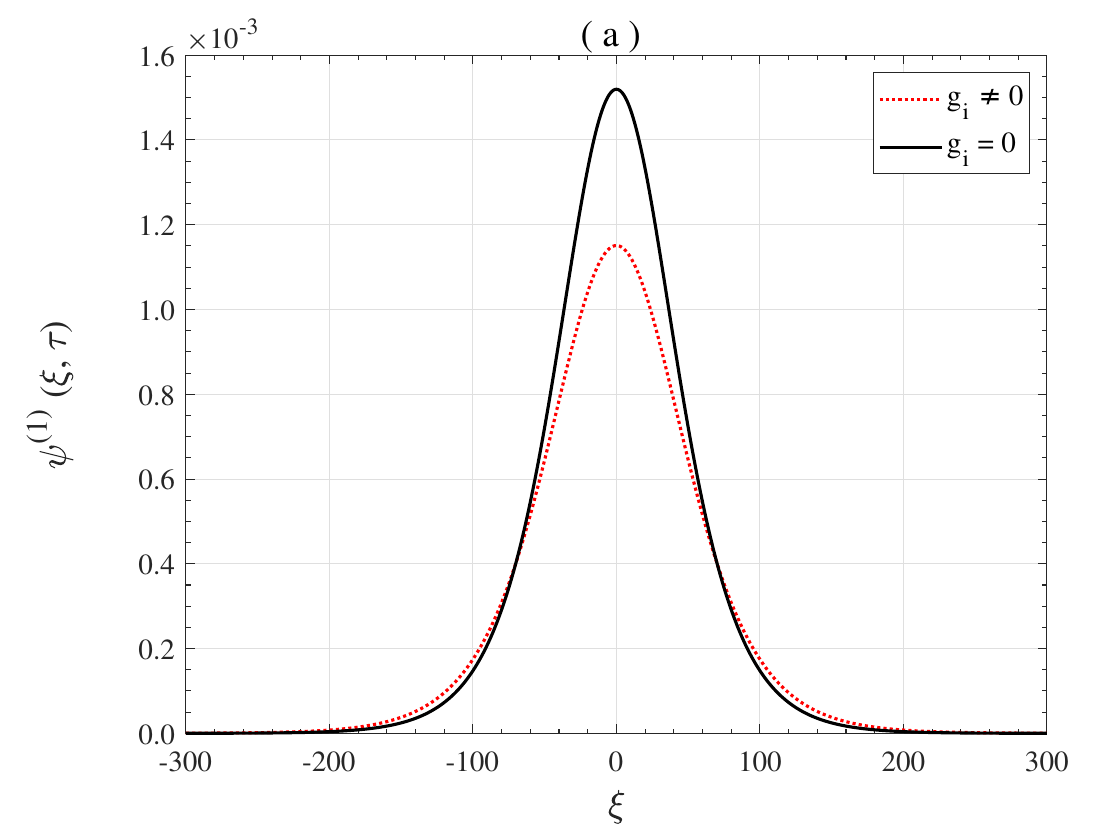}
\quad
\includegraphics[width=8.5cm]{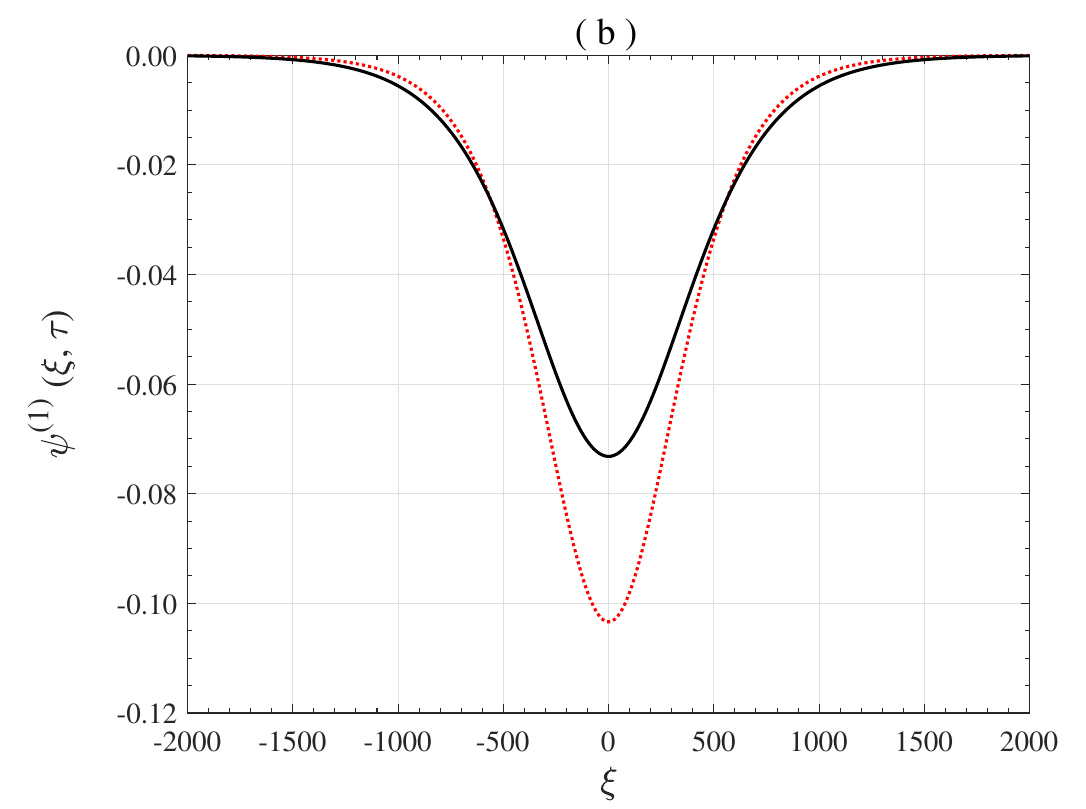}
\caption{Profiles of DIA solitons [Eq. \eqref{eqn_time_dependence_analytical}] are shown in the presence [Subplot (a)] and absence [Subplot (b)] of positive ion-beam density for two different cases: presence ($g_{\textrm{i}} = 0.5$) and absence ($g_{\textrm{i}} = 0$) of ion creation in plasmas.}
\label{fig_11}
\end{figure}
\begin{figure}[!h]
\centering
\includegraphics[width=8.5cm]{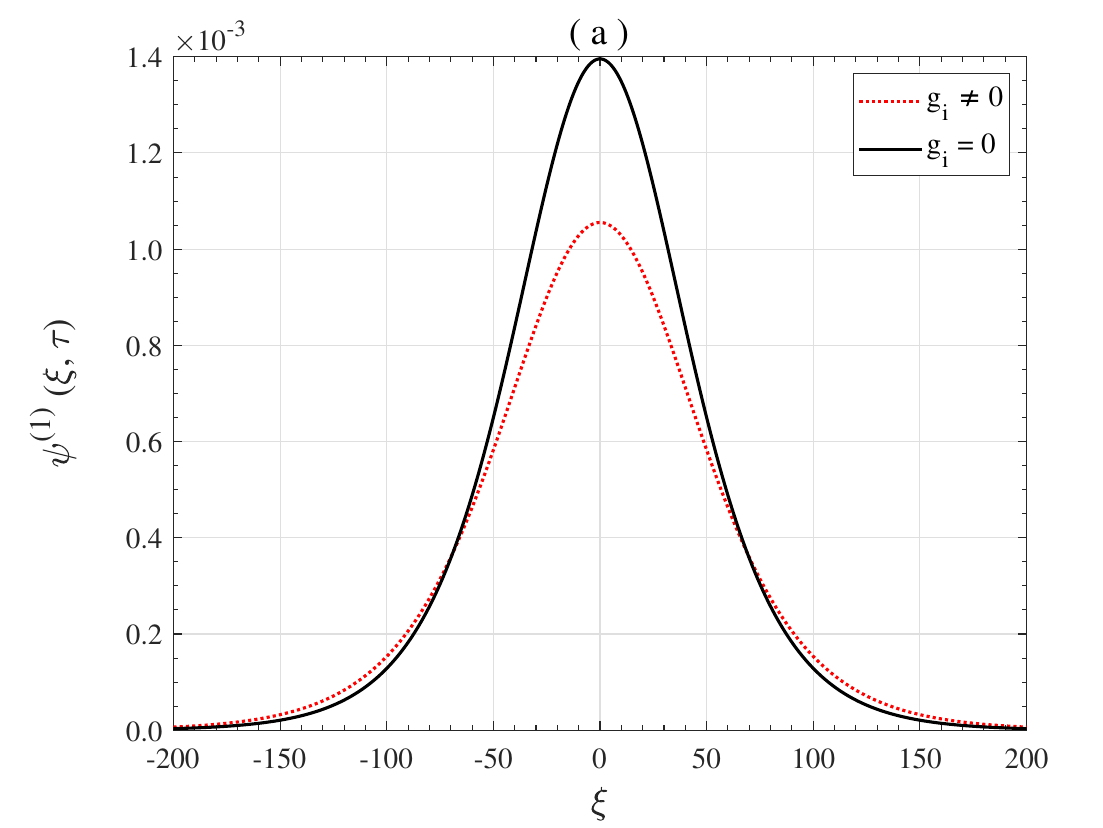}
\quad
\includegraphics[width=8.5cm]{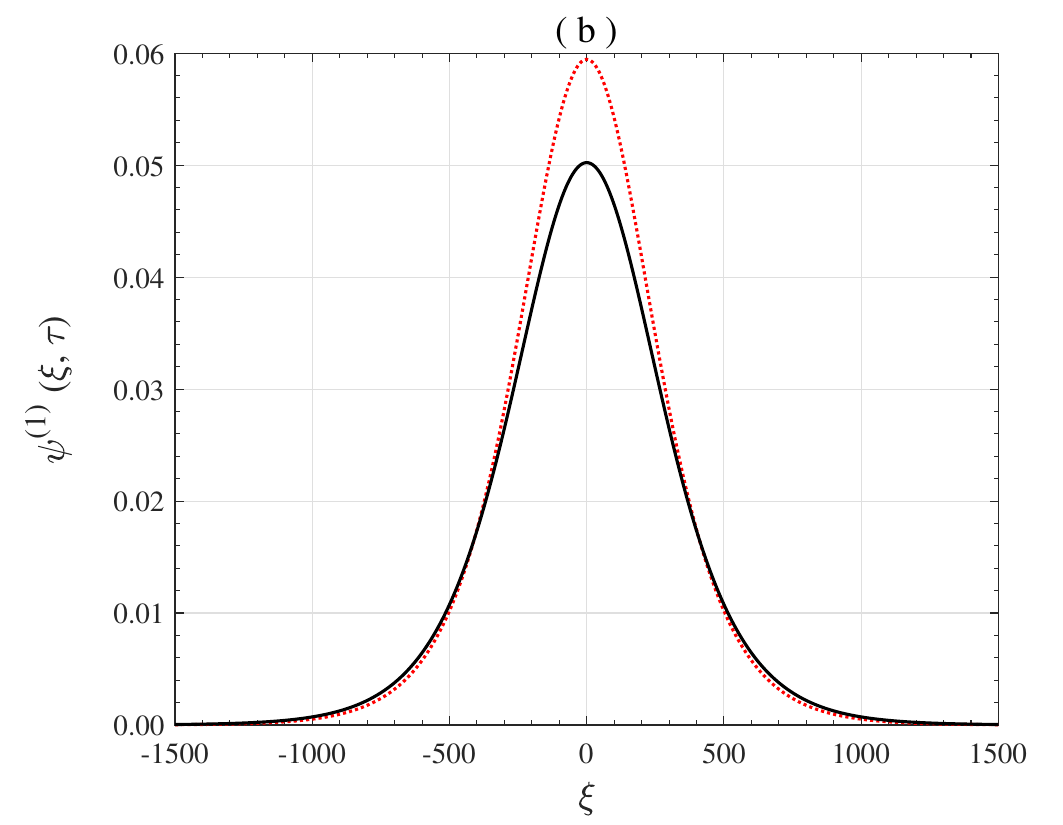}
\caption{Profiles of DIA solitons [Eq. (\ref{eqn_time_dependence_analytical_adiabatic})] are shown in the presence [Subplot (a)] and absence [Subplot (b)] of positive ion-beam density for two different cases: presence ($g_{\textrm{i}} = 0.5$) and absence ($g_{\textrm{i}} = 0$) of ion creation in plasmas.}
\label{fig_12}
\end{figure}
\begin{figure}[!h]
\centering
\includegraphics[width=8.5cm]{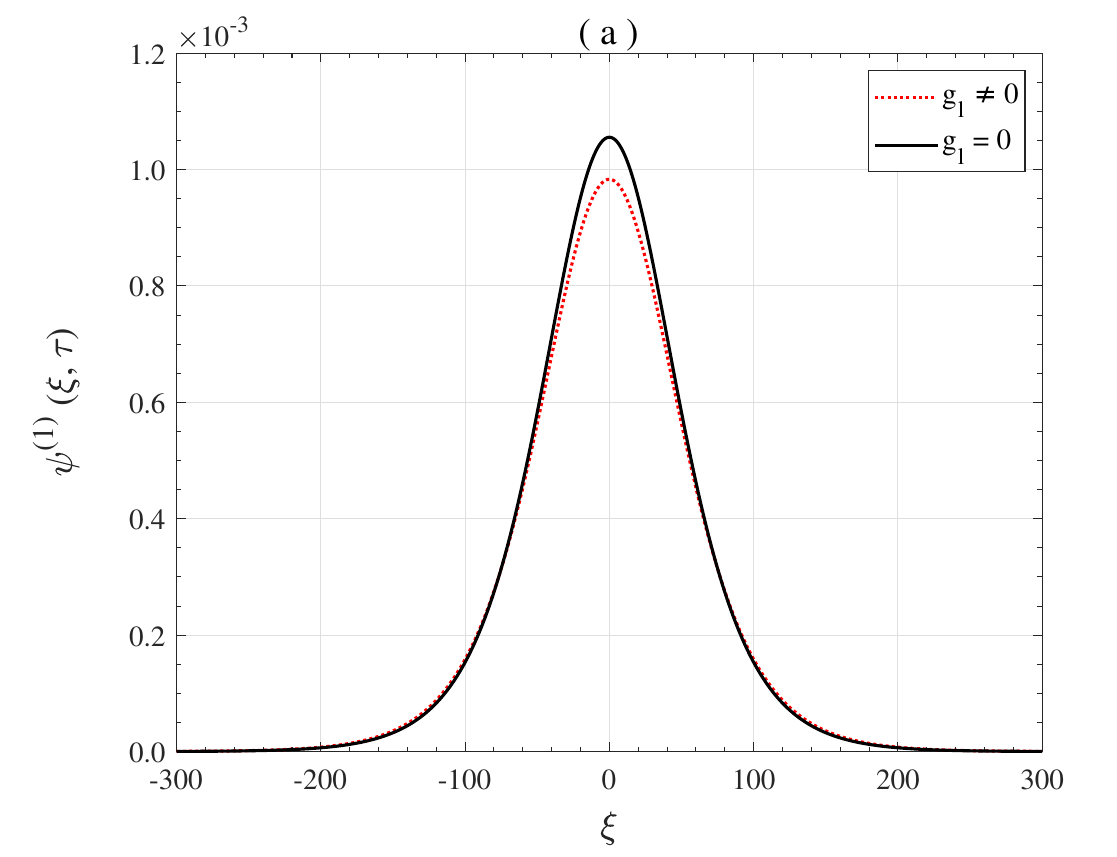}
\quad
\includegraphics[width=8.5cm]{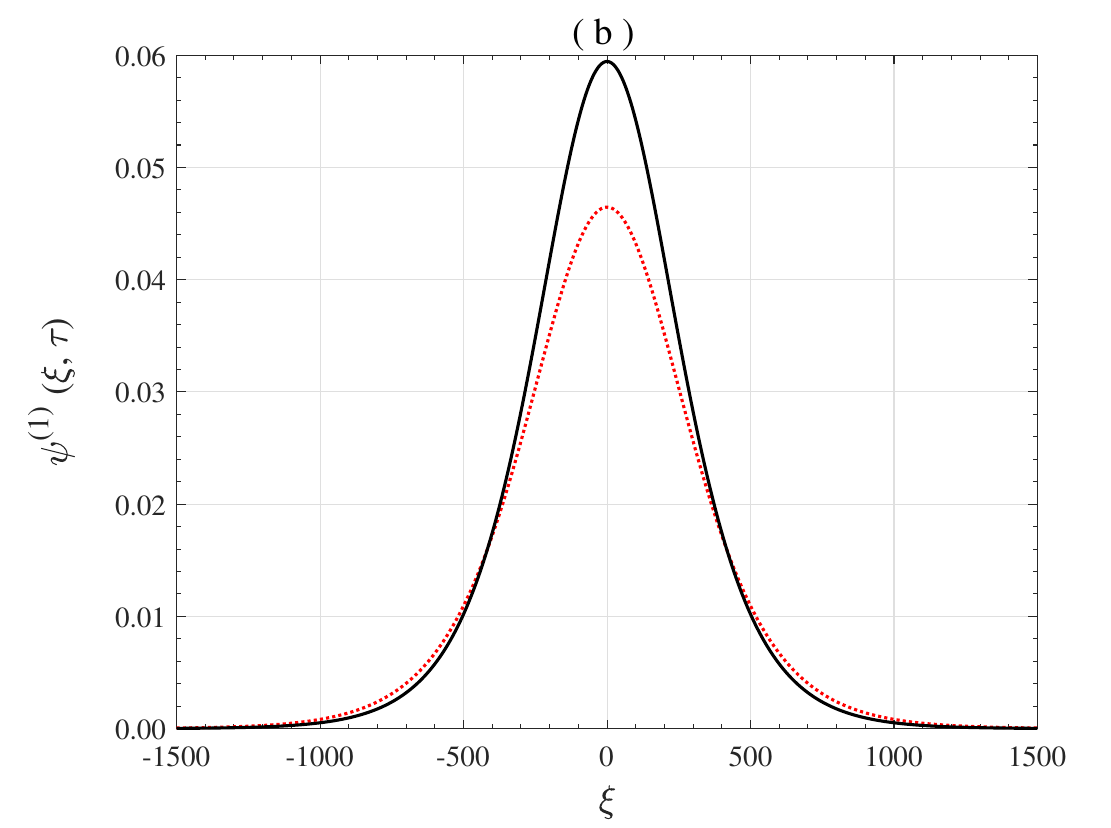}
\caption{Profiles of DIA solitons [Eq. (\ref{eqn_time_dependence_analytical_adiabatic})] are shown in the presence [Subplot (a)] and absence [Subplot (b)] of the positive ion-beam density. The dotted and solid lines, respectively, correspond to the cases when the influence of ion loss is included ($g_{\textrm{l}} = 0.5$) and when there is no ion loss  $g_{\textrm{l}} = 0$).}
\label{fig_13}
\end{figure}
\begin{figure}[!h]
\centering
\includegraphics[width=8.5cm]{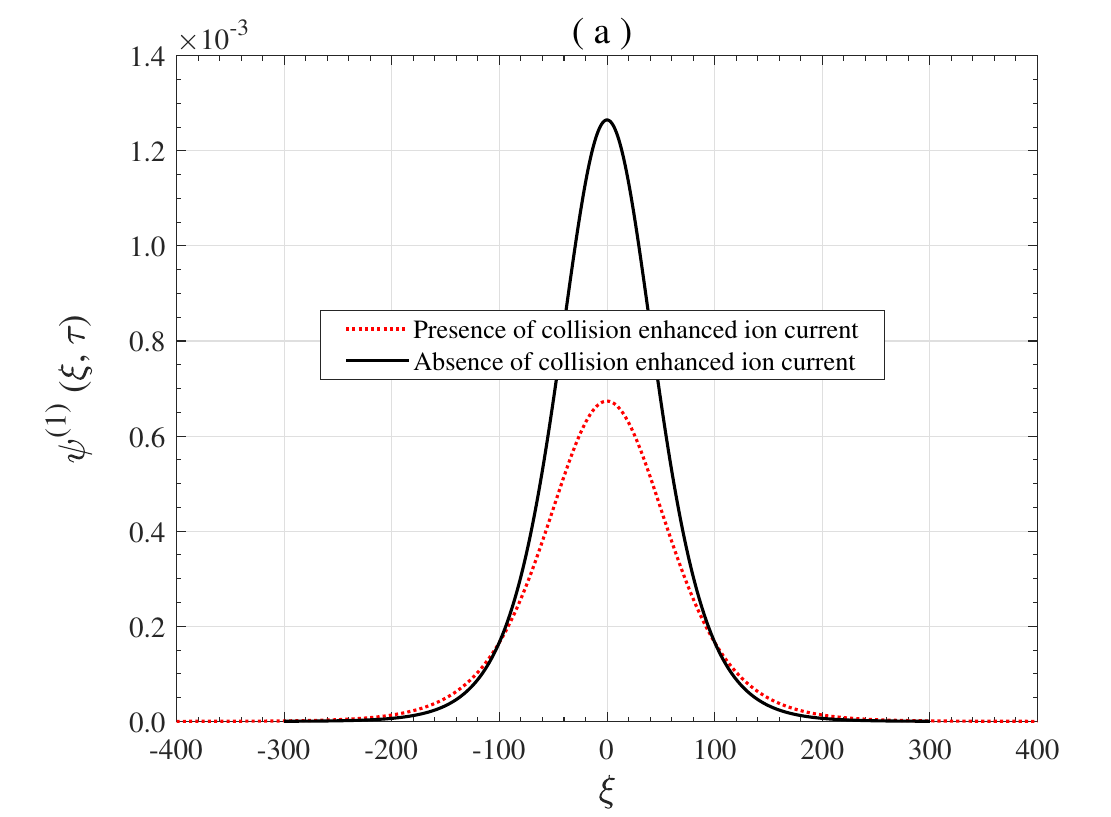}
\quad
\includegraphics[width=8.5cm]{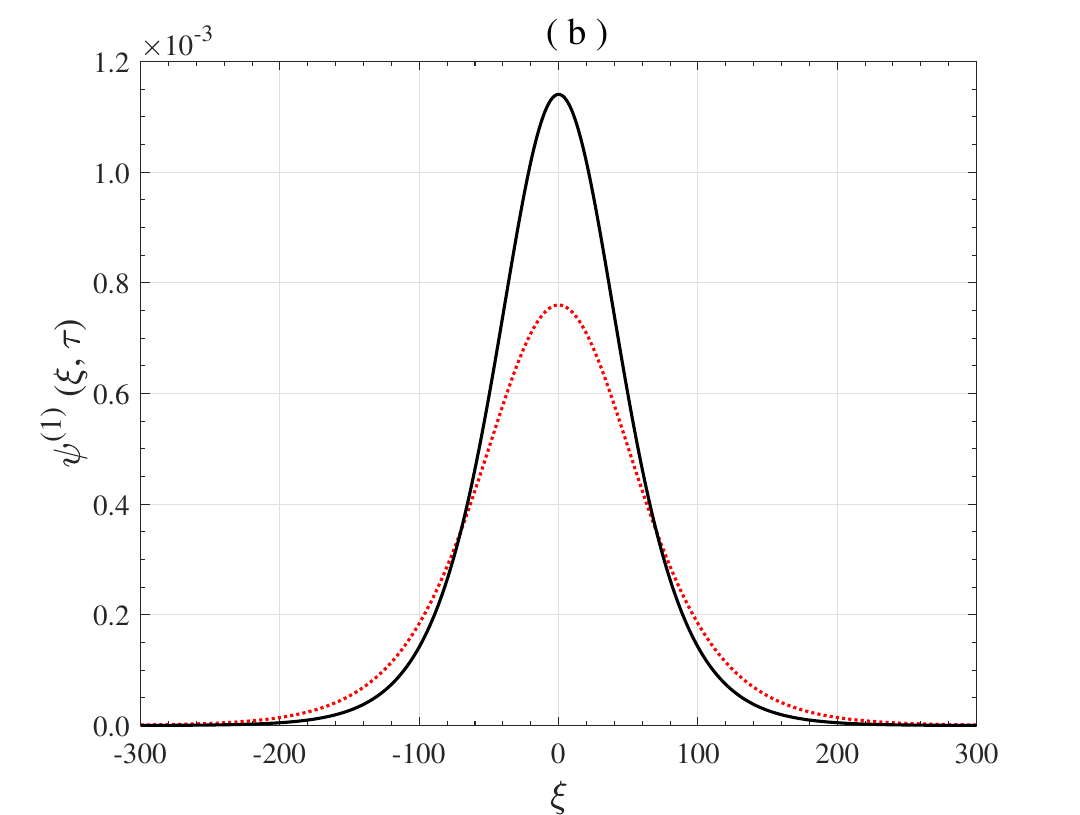}
\caption{Profiles of DIA solitons are shown in the absence (solid lines) and presence (dotted lines) of collision enhanced ion current. Subplots (a) and (b) correspond to Eqs. \eqref{eqn_time_dependence_analytical} and (\ref{eqn_time_dependence_analytical_adiabatic}) respectively.}
\label{fig_14}
\end{figure}

We numerically analyze the characteristics of DIA solitons with the variations of the parameters: $\delta_b$, $\omega_i$, $\theta$, $u_{b0}$, and $\sigma_b$ and exhibit them graphically in Figs. \ref{fig_5}--\ref{fig_9}. We also study the influences of dust charge fluctuations, ion creation, ion loss, and collision enhanced ion current on the profiles of DIA solitons and exhibit the results in Figs. \ref{fig_10}-\ref{fig_14}. Before we analyze the characteristics of DIA solitons, it is important to note that their profiles typically depend on the nonlinear ($P$ or $P_1$), dispersion ($R$ or $R_1$), and the damping ($S$ or $S_1$) coefficients. While the soliton amplitude is influenced by wave nonlinearity and varies inversely with it, the wave dispersion directly influences the width of solitons. On the other hand, the dissipation ($S$ or $S_1$) contributes only to the soliton speed, $U(\tau)$ (which directly influences the soliton amplitude but inversely to the width), causing decay with time and hence decay of the soliton amplitude. However, we consider a fixed time to exhibit the profiles. So, we can not observe the decay of soliton amplitude with time. However, due to variations of $S$ or $S_1$ by the effects of the ion creation and ion loss, ion-neutral collision, and dust charge fluctuations, both amplitude and width of solitons will be altered.  Furthermore, the polarity of solitons, i.e., whether they are compressive (with positive potential) or rarefactive (with negative potential) typically depends on the nonlinearity ($P$ or $P_1$). We find that in the case of nonadiabatic dust charge variation, we have $P>0$ ($P<0$) in the presence (absence) of a positive ion-beam. However, the nonlinear coefficient, $P_1$ is always positive in plasmas with adiabatic dust charge variations. Thus, it follows that while both compressive and rarefactive DIA solitons can exist in ion-beam-driven dusty plasmas with nonadiabatic dust charge variations, ion-beam-driven dusty plasmas with adiabatic dust charge variations can support only compressive solitons. In the following, we will study the characteristics of DIA solitons with the variations of the parameters stated before in more detail. Here, one important point is that when the nonlinear coefficient ($P$ or $P_1$) of the damped KdV equation [\eqref{eqn_kdv_modified} or \eqref{eqn_kdv_modified_adiabatic}] vanishes, the equation fails to describe the evolution of DIA solitons. In such cases, we obtain a modified KdV equation with higher-order corrections of perturbations but it is beyond the scope of the present study.    
\par
 Figure \ref{fig_5} shows the profiles of DIA solitons (at a fixed time $\tau$) with the variations of the beam to ion density ratio $(\delta_b)$ in the cases of nonadiabatic [Subplot (a)] and adiabatic [Subplot (b)] dust charge variations. On increasing the concentration of positive ion beam density, the negativity of dust charge increases as the streaming velocity of the positive ion beam is below the critical beam velocity (approximately $46$ times the ion-acoustic speed). We note that the dust charge number inversely varies with the phase velocity, which can be verified by using the quasineutrality condition in Eq. (\ref{eqn_phase_velocity_weakly}) for nonadiabatic dust charge variations and in Eq. (\ref{eqn_phase_velocity_adiabatic}) for adiabatic dust charge variations. Thus, on increasing the number density of the positive ion beam from $0.1$ to $0.6$, the phase velocity of DIA solitary waves decreases, resulting in an increase in the nonlinear coefficient $P$ (or $P_{1}$) and a decrease in the dispersion coefficient $R$ (or $R_{1}$). Thus, both solitary amplitude and width decrease with increasing the concentration of positive ion beams in a positive ion beam-driven dusty plasma. When we increase $\delta_b$ from $0.1$ to $0.6$, the amplitude and width decrease from $0.0019$ to $0.0010$ and from $83.03$ to $62.70$, respectively in the case of nonadiabatic dust charge variations  [Subplot (a)], and from $0.002$ to $0.0009$ and from $77.31$ to $61.97$ in the case of adiabatic dust charge variations [Subplot (b)]. We also observe that the qualitative features of DIA solitons in plasmas with nonadiabatic and adiabatic dust charge variations remain the same except for the magnitudes of amplitudes and widths. Since we observe reductions in the amplitudes and widths of solitons by the influence of the ion-beam density, DIA solitons in beam-driven dusty plasmas would evolve with lower energies than plasmas without ion beam.  
 %%%%%%%%%%%%%%%%%%%%%%%%%%%%%%%%%%%%%%%
%%%%%%%%%%%%%%%%%%
\par 
The influences of the external magnetic field on the profiles of DIA solitons are shown in Fig. \ref{fig_6}. Since the magnetic field only influences the dispersion coefficient ($R$ or $R_1$) via the ion gyrofrequency $(\omega_i)$ inversely, the soliton amplitude remains unchanged. Still, the width gets significantly reduced as the values of $\omega_i$ slightly increase. We observe reductions of widths from $313.39$ to $62.69$ in the case of nonadiabatic dust charge variations [Subplot (a)] and from $308.80$ to $61.76$ in the case of adiabatic dust charge variations [Subplot (b)] as $\omega_i$ increases from $0.01$ to $0.05$. Thus, DIA solitons evolve with higher energies in weakly magnetized dusty plasmas than plasmas with strong magnetic fields.  
\par 
The obliqueness of wave propagation about the magnetic field $(\theta)$ also significantly influences the characteristics of DIA solitons, as shown in Fig. \ref{fig_7}. Having known the dependency of all the coefficients of the KdV equation on $\theta$, we find that the soliton amplitude gets greatly reduced and width significantly enhanced as the angle increases in the interval $0<\theta<\pi/2$. 
We find that as $\theta$ increases from $10^{\circ}$ to $60^{\circ}$, the soliton amplitude decreases from $0.0014$ to $0.0003$, and the width increases from $19.58$ to $208.04$ in the case of nonadiabatic dust charge variations [See subplot (a)]. However, when we consider the adiabaticity of dust charge variations, the soliton amplitude decreases from $0.0013$ to $0.0003$, and the width increases from $19.29$ to $205.28$ [See subplot (b)] with the same variations of $\theta$. We also note that given a fixed value of $\theta$, the relative magnitudes of soliton amplitude and width can be higher in the case of adiabatic dust charge variation than in the nonadiabatic case. Furthermore, the solitons become narrower and taller as the wave propagation direction approaches the magnetic field, and they become wider and shorter at an angle nearly perpendicular to the magnetic field. In both the limiting cases of wave propagation, the DIA solitons evolve with higher energies at a fixed time. 
%%%%%%%%%%%%%%%%%%%%%%% 
%%%%%%%%%%%%%%%%%%%%%%%%%%%%%%%%
\par
Figure \ref{fig_8} shows the profiles of DIA solitons by the influence of the ion-beam streaming velocity in the cases of nonadiabatic [Subplot (a)] and adiabatic [Subplot (b)] dust charge variations. We note that of the coefficients $P$, $R$, and $S$, only the dissipation coefficient  $S$ that signifies the order of decay of the soliton velocity and hence a decrease (increase) in the amplitude (width) depends on the beam streaming velocity $u_{b0}$ via $\chi_1$ (associated with the dust charge fluctuation frequency $\nu_{\rm{ch}}$). However, changes in the soliton amplitude and width with the variations of $u_{b0}$ are insignificant in the case of nonadiabatic dust charge variation. On the other hand, when we consider the adiabaticity of dust charge variation, the influence of the beam streaming velocity on the soliton profiles becomes noticeable. We observe that the amplitude and width get reduced as one increases the beam streaming velocity. As the beam velocity increases, the dust charge number decreases, which leads to an increment of the nonlinear coefficient $P_{1}$ and a decrement of the dispersion coefficient $R_{1}$, and hence a decrease in both amplitudes and widths of solitons. Thus, the streaming of ion beams can be significant in dusty plasmas with adiabatic dust charge variations, and higher its magnitude than the ion-acoustic speed lower the soliton energy.
%%%%%%%%%%%%%%%%%%%%%%%%%%%%%%%%%%%%%%
%%%%%%%%%%%%%%%
\par
We also study the influence of the beam temperature on the characteristics of DIA solitons and exhibit the results in Fig. \ref{fig_9}. Here, we consider values of the beam temperature so that the beam thermal velocity remains lower than the beam streaming velocity. We observe that in the cases of nonadiabatic [Subplot (a)] and adiabatic [Subplot (b)] dust charge variations, an enhancement of the beam to ion temperature ratio ($\sigma_b$), leads to a significant enhancement of both the soliton amplitude and width.
Since the beam temperature directly influences the phase velocity of DIA waves [Eqs. (\ref{eqn_phase_velocity_weakly}) and (\ref{eqn_phase_velocity_adiabatic})], the nonlinear coefficient $P$ (or $P_{1}$) decreases with an increase in the phase velocity. Also, the nonlinear coefficients $P$ (or $P_{1}$) have an inverse relation with soliton amplitude [Eqs. (\ref{eqn_solitary_amplitude}) and (\ref{eqn_solitary_amplitude_adiabatic})]. Thus, the soliton amplitude increases with an increase in the beam temperature. In addition, the dispersion coefficient $R$ (or $R_{1}$) has a direct relation with the beam temperature and the soliton width [Eqs. (\ref{eqn_solitary_width}) and (\ref{eqn_solitary_width_adiabatic})]. As a result, the soliton width also increases with an enhancement in the beam temperature. For example, if we increase the ratio $\sigma_b$ from $0.004$ to $0.02$, the soliton amplitude increases from $0.00029$ to $0.00115$ and from $0.00026$ to $0.00106$ in the cases of nonadiabatic and adiabatic dust charge variations respectively. The corresponding increments of the soliton widths are from $56.72$ to $62.69$ and from $56.53$ to $61.76$. Thus, the beam thermal energy plays a crucial role in the evolution of DIA solitons with higher energies.
%%%%%%%%%%%%%%%%%%%%%%%%%%%%%%%%%%%%%%%%%%%
\par 
The presence of ion-beam, dust charge fluctuations, ion creation, ion loss, and ion-neutral and ion-dust collisions can also significantly influence the characteristics of DIA solitons, as shown in Figs. \ref{fig_10}-\ref{fig_14}. Figure \ref{fig_10} shows the soliton profiles in the presence [Subplot (a)] and absence [Subplot (b)] of the ion beam by the effects of constant dust charge and dust charge fluctuations. We observe that as the beam density is zero, or we disregard the beam species from plasmas, the soliton polarity changes from compressive to rarefactive type. Also, the effect of the dust charge fluctuation is to enhance the soliton amplitude (the width remains almost unchanged) in beam-driven dusty plasmas [Subplot (a)]. In contrast, the effect is more pronounced in the rarefactive solitons, i.e., DIA solitons in beam-free dusty plasmas with a significant decrement (in magnitude) of the wave amplitude. We also note similar effects on the compressive and rarefactive solitons in the presence and absence of ion creation in plasmas. 
We study the influence of ion creation on the profiles of DIA solitons in the presence (solid lines) and absence (dotted lines) of positive ion-beam density in dusty plasmas with nonadiabatic (\ref{fig_11}) and adiabatic (\ref{fig_12}) dust charge variations. We observe that while the amplitudes (in magnitudes) of compressive solitons [Figs. \ref{fig_11}(a) and \ref{fig_12}(a)] get reduced by the effects of ion creation in ion-bean-driven dusty plasmas with nonadiabatic and adiabatic dust charge variations, the same get significantly enhanced in plasmas without ion beam [Figs. \ref{fig_11}(b) and \ref{fig_12}(b)]. Here, one should note the existence of rarefactive DIA solitons in dusty plasmas without ion beam and with nonadiabatic dust charge variations, and ion creation does not significantly influence the soliton width. Thus, DIA solitons in ion-beam-driven dusty plasmas with dust charge fluctuations propagate with lower energies by the influence of ion creation.   
%%%%%%%%%%%%%%%%%%%%%%%5
\par
Another important factor that can influence the characteristics of wave propagation is the ion loss (sink) term. However, it does not influence DIA solitons when we consider the nonadiabatic dust charge variation in beam-driven dusty plasmas.  So, we show the results in Fig. \ref{fig_13} for DIA solitons only in the case of adiabatic dust charge variations with [Subplot (a)] and without [Subplot (b)] ion beam. Typically, ion loss influences the ion dynamics and thereby modifies the wave dispersion and damping. While the ion loss causes soliton amplitude to grow in ion-beam-driven plasmas, the same gets significantly reduced without the ion beam. Typically, in the presence of the ion loss term in beam-driven plasmas, the soliton Mach number increases, and the soliton amplitude increases as well. However, the Mach number decreases in the absence of the positive ion beam, which results in a decrement of soliton amplitude in the absence of the ion beam. Similar to the effects of ion creation, the ion loss does not significantly influence the soliton amplitude.
On the other hand, the ion-neutral collision significantly influences the dust charging process by increasing the ion flux to the surface of dust grains and thus reduces the dust charge number. Such a reduction in the dust charge also affects the solitary wave structures. We see that in both the cases of nonadiabatic [Subplot (a)] and adiabatic [Subplot (b)] dust charge variations, the soliton amplitude gets significantly reduced (and the width enhanced) by the influence of the ion-neutral collision enhanced current (See Fig. \ref{fig_14}). For example, in the case of nonadiabatic dust charge variation, the soliton amplitude reduces from $0.0013$ to 0.$0007$ while the width increases from $59.80$ to $75.50$ [Subplot (a)]. Also, we observe a reduction of the soliton amplitude from $0.0011$ to $0.0007$ and an enhancement of the width from $58.76$ to $74.98$ in the case of adiabatic dust charge variations [Subplot (b)]. It follows that the soliton energy greatly reduces and the wave gets damped more quickly in account of the ion-neutral collision enhanced current in dusty plasmas with dust charge fluctuations.
%%%%%%%%%%%%%%% 
\vspace{0.2cm}
\section{SUMMARY}\label{sec-summ}
We have studied the nonlinear evolution of small-amplitude DIA solitary waves in a positive ion-beam-driven collisional dusty magnetoplasma with dust charge fluctuations. Using the reductive perturbation technique, we have derived the evolution equations for DIA solitary waves in the form of KdV equations and studied the characteristics of DIA solitons in two limiting cases of nonadiabatic and adiabatic dust charge fluctuations. We have shown that the dust charge number gets significantly modified by the ion beam current while contributing to the electron, ion, and collision enhanced currents, and the ion creation and loss, and ion-neutral and ion-dust collisions lead to damping of DIA solitary waves. In addition, the contribution from the dust charge fluctuation to the wave damping is also noticeable when we consider the adiabaticity of the dust charge fluctuation. Furthermore, we have observed that only compressive type (with positive potential) DIA solitons exist in beam-driven dusty plasmas with nonadiabatic and adiabatic dust charge variations. However, DIA solitons of the rarefactive-type (with negative potential) may exist in dusty plasmas without the influence of the ion beam. In the following, we summarize our main findings from the study as follows:
\begin{itemize}
\item The ion-neutral collision enhanced current reduces the equilibrium dust charge number by an increase in the ion flux flowing into the dust-grain surface. However, with increasing beam density, the same increases. The streaming velocity of the positive ion beam also significantly influences the evolution of the dust charge.
\item The linear phase velocity of DIA waves in plasmas with adiabatic dust charge variations appears to somewhat generalize the same in plasmas with nonadiabatic dust charge variations. Furthermore, the phase velocity in the case of nonadiabatic dust charge variation is independent of the dust number density and thus agrees with ion-acoustic waves in ion-beam-driven plasmas without charged dust. 
The phase velocity tends to decay with increasing values of the beam density, and the decreasing rate becomes faster with the higher beam streaming velocity.
\item The soliton energy decays with time, and the decay rate can become faster the larger the beam streaming velocity. The effect is more noticeable in the case of an adiabatic dust charge variation than in the nonadiabatic one.  Furthermore, the beam density significantly influences the soliton profiles in reducing both the amplitude and width and the soliton energy. However, while the beam streaming velocity has no significant influence on the soliton profiles in the case of nonadiabatic dust charge variation, it can reduce both the amplitude and width when we consider the adiabatic dust charge variation. In both these cases, the soliton amplitude and width get significantly enhanced by the influence of higher beam temperatures.
\item The static magnetic field does not influence the soliton amplitude but reduces the width by increasing its strength. Also, the obliqueness $(\theta)$ of wave propagation about the magnetic field significantly reduces the soliton amplitude but increases the width as $\theta$ tends to enhance in $0<\theta<\pi/2$.
\item In beam-driven dusty plasmas, while the ion creation reduces the DIA soliton amplitude, the ion loss enhances the amplitude without noticeable influence on the width. Also, the soliton amplitude gets significantly reduced, and the width gets enhanced by the influence of the collision enhanced ion current.
\end{itemize}
To conclude, the present work systematically explores the characteristics of dust-ion acoustic solitons in collisional (weakly ionized) magnetized positive ion-beam-driven dusty plasmas with the physical parameters that are relevant to laboratory plasmas. The results should help understand the energy transport phenomena as well as the stability of solitons in dusty plasmas that are crucial in industrial environments with technological applications in diverse fields.    
\section*{Acknowledgments}
Author N. P. Acharya would like to acknowledge the University Grants Commission, Bhaktapur, Nepal for the Ph.D. Fellowship (PhD-78/79-S\&T-17). \\
(S. Basnet) This research was supported by an appointment to the Young Scientist Training (YST) program at the APCTP through the Science and Technology Promotion Fund and Lottery Fund of the Korean Government. This was also supported by the Korean Local Governments-Gyeongsangbuk-do Province and Pohang City.
\ifCLASSOPTIONcaptionsoff
  \newpage
\fi

\begin{IEEEbiography}[{\includegraphics[width=1in,height=1.25in, clip,keepaspectratio]{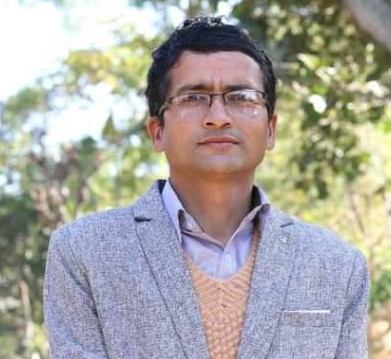}}]{Num Prasad Acharya was born in Pyuthan, Bijuli-02, Nepal, in 28 September 1987. He received the B. Sc. and M. Sc. Degree in Physics from Tribhuvan University, Nepal. He is currently pursuing a Ph.D. degree in plasma physics from the Central Department of Physics, Tribhuvan University, Kirtipur, Nepal. He is a member of Association of Asia Pacific Physical Societies-Division of Plasma Physics (AAPPS-DPP).\\
The research interest of Mr. Acharya is linear and non-linear wave instabilities in the ordinary and dusty plasmas.}
\end{IEEEbiography}
\begin{IEEEbiography}[{\includegraphics[width=1in,height=1.25in,clip,keepaspectratio]{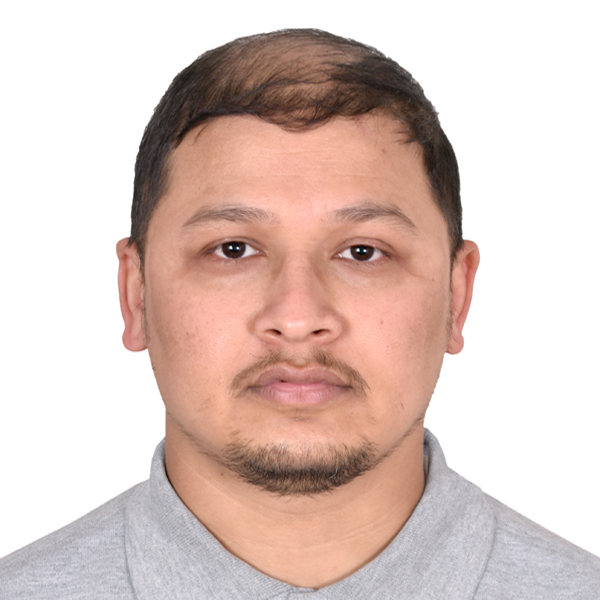}}]{Suresh Basnet was born in Itahara-02, Morang, Nepal, in 14 August 1986. He received the B. Sc., M. Sc. and Ph.D. Degree in Physics from Tribhuvan University, Kirtipur, Nepal. He was a Postdoctoral researcher at Central Department of Physics, Tribhuvan University, Nepal from 2023 to 2024. Currently, he is a Postdoctoral fellow at the Asia Pacific Center for Theoretical Physics, Pohang, Republic of Korea. He is a member of Plasma Science Society of India (PSSI) and Association of Asia Pacific Physical Societies-Division of Plasma Physics (AAPPS-DPP).\\
His research interests are plasma-wall transition, linear and non-linear waves in dusty plasma, space, and astrophysical magnetized plasma features. He is currently working on the properties of magnetized plasma.}
\end{IEEEbiography}
\begin{IEEEbiography}[{\includegraphics[width=1in,height=1.25in,clip,keepaspectratio]{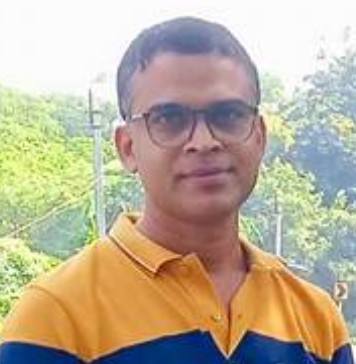}}]{Amar Prasad Misra received the B.Sc. degree in Mathematics and M.Sc. degree in Applied Mathematics from the University of Calcutta, India, and the Ph.D. degree from Jadavpur University, India. He was a postdoctoral fellow (2009-2011) in the Department of Physics, Umeå University, Sweden. He has been a faculty member in the Department of Mathematics, Visva Bharati University, India, since 2005, where he serves on the Research Board. He is a life member of the Plasma Science Society of India (PSSI) and the Association of Asia Pacific Physical Societies-Division of Plasma Physics (AAPPS-DPP). He has published over 130 research papers in international peer-reviewed journals. He is currently Editor-in-Chief of Physica Scripta, Editorial Board member of Scientific Reports, and Associate Editor for Low-Temperature Plasma Physics at Frontiers in Physics and Frontiers in Astronomy and Space Sciences. His current research interests include nonlinear waves and instabilities in plasmas, atmospheric waves, magnetohydrodynamics, chaos in nonlinear dynamical systems, chaos-based cryptography, and network security.}
\end{IEEEbiography}
\begin{IEEEbiography}
[{\includegraphics[width=1in,height=1.25in,clip,keepaspectratio]{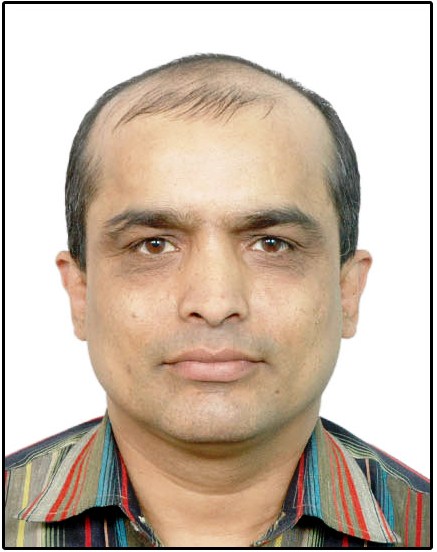}}]{Raju Khanal was born in Tehrathum, Nepal, in 1969. He earned his M.Sc. degree in physics from Tribhuvan University, Kirtipur, Nepal, in 1994, and his Ph.D. in plasma physics from the Universität Innsbruck, Austria, in 2003. Since 1997, he has been affiliated with Tribhuvan University, where he has been a Professor of Physics since 2015.}
\end{IEEEbiography}
\end{document}